\documentclass[preprint,amsmath,amssymb,showpacs, aps,longbibliography]{revtex4-1}

\usepackage[dvips]{graphicx}

\begin{document}

\title{Phase diagrams and free-energy landscapes
for model spin-crossover materials with antiferromagnetic-like
nearest-neighbor and ferromagnetic-like long-range interactions}

\author{C.~H. Chan}
\email{seahoi2001@gmail.com }

\author{G. Brown}
\email{gbrown@fsu.edu}

\author{P.~A. Rikvold}
\email{prikvold@fsu.edu}

\affiliation{Department of Physics, Florida State University, Tallahassee, Florida 32306-4350, USA}

\date{\today }

\begin{abstract}
We present phase diagrams, free-energy landscapes, and order-parameter distributions for a model spin-crossover material with a two-step transition between the high-spin and low-spin states (a square-lattice Ising model with antiferromagnetic-like nearest-neighbor and ferromagnetic-like
long-range interactions) [P.~A.\ Rikvold {\it et al.\/}, Phys.\ Rev.\ B {\bf 93}, 064109 (2016)]. The results are obtained by a recently introduced, macroscopically constrained Wang-Landau Monte Carlo simulation method [C.~H.\ Chan, G.\ Brown,
and P.~A.\ Rikvold, Phys.\ Rev.\ E {\bf 95}, 053302 (2017)]. The method's computational efficiency enables calculation of thermodynamic quantities for a wide range of temperatures, applied fields, and long-range interaction strengths.
For long-range interactions of intermediate strength, tricritical points in the phase diagrams are replaced by pairs of critical end points and mean-field critical points that give rise to horn-shaped regions of metastability. The corresponding free-energy landscapes offer insights into the nature of asymmetric, multiple hysteresis loops that have been experimentally observed in spin-crossover materials characterized by competing short-range interactions and long-range elastic interactions.
\end{abstract}

\maketitle

\section{introduction}\label{sec:Introduction}

Spin-crossover (SC) materials are molecular crystals in which the individual molecules contain
transition metal ions that can exist
in two different spin states: a low-spin ground state (LS) and
a high-spin excited state (HS). Molecules in the HS state have
larger volume and higher effective degeneracy than those in the LS state
\cite{PhysRevB.77.014105,PhysRevB.84.054433,NAKA12,ENAC05,PhysRevLett.98.247203,PhysRevLett.100.067206,BOUS11,HALC13}.
Due to its higher degeneracy, crystals of such
molecules can be brought into a majority excited HS
state by increasing temperature, changing pressure or magnetic field,
electrochemical stimuli, or exposure to light
\cite{PhysRevB.80.064414,PhysRevLett.100.067206,PhysRevB.82.020409,ENAC05,GUTL94,CHON10,OHKO11,ASAH12,CHAK14,TOKO16,TOKO17}.
The size difference between the HS and LS molecules causes local elastic distortions
that lead to effective long-range elastic interactions mediated by the
macroscopic strain field \cite{PhysRevLett.98.247203,TEOD82,ZHUX05}.
In addition to such long-range interactions, the materials will also typically have local
interactions caused by, e.g., quantum-mechanical exchange or geometric restrictions.
These intermolecular interactions may cause first-order phase transitions that can render the
HS state metastable and lead to hysteresis when exposed to time-varying fields
\cite{GUTL94,Seiji_tokyo}.
In the case of optical excitation into the metastable phase, this phenomenon is known as
light-induced excited spin trapping (LIESST) \cite{GUTL94,PhysRevB.80.064414,TOKO17}.
The metastable properties in combination with the SC materials' sensitivity to a wide range of
external stimuli make them promising candidates for applications such as
switches, displays, memory devices, sensors,
and actuators \cite{OHKO11,CHAK14,KAHN98,LINA12}.

In the SC literature, the phase transitions caused by the short-range and long-range
interactions are often discussed using an Ising-like pseudospin formulation,
in which the HS state is represented as $s=+1$ and the LS state as $s=-1$.
This is the representation we will use in this paper.
It has the advantage of a high degree of symmetry, and it enables easy
reference to studies of other Ising-like models.
To minimize the strain energy,
the elastic long-range interaction favors different molecules being in the
same state (LS-LS or HS-HS). In this pseudospin language it is therefore called
ferromagnetic-like, or simply ferromagnetic.
The short-range interactions depend on the particular material
and may either be ferromagnetic-like, or they may favor neighboring molecules in opposite states
(LS-HS), which is analogously called antiferromagnetic-like, or simply antiferromagnetic.
We emphasize that this nomenclature only represents
an analogy and {\em does not\/} imply a magnetic origin of the
interactions. In the remainder of this paper, we will use the simplified terms,
ferromagnetic and antiferromagnetic,
for interactions that favor uniform and checkerboard spin-state arrangements, respectively.

If the short-range interaction is ferromagnetic, it has been found that
adding even a very weak long-range interaction causes the universality class of the critical point
to change from the Ising class to the mean-field class \cite{PhysRevB.84.054433,NAKA12}.
On the other hand, if the short-range interaction is antiferromagnetic, the critical line
will terminate at a certain point, with the appearance of metastable regions in the phase diagram,
bounded by sharp spinodal lines \cite{BROWN201420,PhysRevB.93.064109}.
Then, with sufficiently strong long-range interaction, new mean-field critical points
emerge in the phase diagrams -- a phenomenon which is not predicted by simple
Bragg-Williams mean-field theory \cite{PhysRevB.93.064109}. These new mean-field critical
points also become the end points for the spinodal lines bounding the metastable regions.

In some SC materials, the transition between the LS and HS phases proceeds as a
two-step transition via an intermediate phase  \cite{KOPP82,ZELE85,PETR87,BOUS92,JAKO92,REAL92,BOIN94,BOLV96,CHER04,BONN08,PILL12,BURO10,LIN12,KLEI14},
giving rise to complex, asymmetrical hysteresis loops.
In the case of Fe(II)[2-picolylamine]$_3$Cl$_2 \cdot$Ethanol \cite{KOPP82},
x-ray diffraction has revealed an intermediate phase, characterized by long-range order on two interpenetrating sublattices with nearest-neighbor
molecules in different states (HS-LS) \cite{CHER03,HUBY04}.
Several of these experimental results were recently reviewed \cite{SHAT15,BROO15}.
This situation can be modeled by an Ising-like model with
antiferromagnetic nearest-neighbor interactions. Various mean-field approximations to this
model have been considered, both without \cite{BOLV96} and with \cite{ZELE85,BOUS92,CHER04}
a long-range ferromagnetic term.

Recently, Rikvold {\em et al.\/} used standard importance-sampling Monte Carlo (MC)
simulations to obtain phase diagrams and hysteresis curves for
such an Ising model with nearest-neighbor
antiferromagnetic interactions and ferromagnetic long-range interactions
approximated by a mean-field equivalent-neighbor (Husimi-Temperley) term
\cite{PhysRevB.93.064109}. (See Hamiltonian in Sec.~\ref{sec:Hamiltonian_Ising_ASFL}.)
To locate the various transition lines in the phase diagram, this method requires separate
simulations for different values of temperature, field, and long-range interaction strength.
This procedure is very computationally intensive,
and phase diagrams could therefore only be drawn for three different interaction strengths.

In the present paper, we  provide detailed phase diagrams for this system with
a range of different long-range interaction strengths, from quite weak to quite strong.
In addition to phase diagrams, we also obtain free-energy landscapes and order-parameter
probability densities in terms of the model's two order parameters,
magnetization ($M$) and staggered magnetization ($M_s$).
To obtain these results with a reasonably modest computational effort,
we use a recently proposed,
macroscopically constrained Wang-Landau (WL) MC algorithm \cite{PhysRevE.94.042125,PhysRevE.95.053302}.
With this method,  a simple analytic transformation of the system energy $E$
enables us to extract results for any combination of temperature, applied field,
and long-range interaction strength from one single, high-precision simulation of the
joint density of states (DOS), $g(E,M,M_{s})$,
for a simple square-lattice Ising antiferromagnet in zero field.
The details of how to use the algorithm to calculate the joint DOS,
and how to extract from it free-energy landscapes, order-parameter probability densities, and
phase diagrams are given in our recent papers, Ref.~\cite{PhysRevE.95.053302,CHAN17conference}.
Here,
we concentrate on the physical aspects of this model SC material and, in particular,
their dependence on the long-range interaction strength.
In the process, we also obtain improved estimates for the positions and shapes of
the first-order coexistence lines in the phase diagrams.

Studies of Ising models with long-range interactions have a long history. Some notable
examples are work on Ising models with weak long-range interactions
by Penrose, Lebowitz, and Hemmer
 \cite{Penrose_Lebowitz_1971,Hemmer_Lebowitz_1976review,Penrose_Lebowitz_book},
and with long-range lattice coupling by Oitmaa and Barber \cite{Oitmaa_Barber_1975}.
Herrero studied small-world networks with both ferromagnetic \cite{PhysRevE.65.066110}
and antiferromagnetic interactions \cite{PhysRevE.77.041102}. Hasnaoui and Piekarewicz \cite{PhysRevC.88.025807} recently used an Ising model with Coulomb long-range interaction to simulate nuclear pasta in neutron stars.
 It should also be mentioned that the Ising model with long-range interactions decaying as $r^{-(d+\sigma)}$ with $d=1,2,3$ and $0<\sigma<d/2$ was studied by Luijten and Bl\"ote
\cite{PhysRevB.56.8945}, and the effect of long-range interactions on phase transitions
in short-range interacting systems were studied by Capel \textit{et al.} \cite{CAPEL1979371}.

The remainder of this paper is organized as follows. In Sec.~\ref{sec:Hamiltonian_Ising_ASFL} we
present the Ising-like model Hamiltonian and its interpretation as a model for SC materials.
In Sec.~\ref{sec:method} we briefly discuss the macroscopically constrained WL algorithm
and present the analytic energy transformation that enables us to extract data for arbitrary
model parameters from a single simulated joint DOS.
We also show how constrained partition functions are
obtained from the joint densities of states, and how the partition functions lead to
free-energy landscapes and order-parameter probability densities.
Sec.~\ref{sec:phase_diagram} contains our main results: phase diagrams, as well as
probability densities and free-energy landscapes at selected phase points.
All these are obtained for several values of the long-range interaction strength, ranging from quite weak to quite strong, and producing a number of topologically different phase diagrams.
Section \ref{sec:conclusion} contains a brief summary and conclusions.
Details of our estimates of finite-size and statistical errors are given in
Appendix \ref{sec:finite_size_error}.

\section{2D Ising-ASFL model}\label{sec:Hamiltonian_Ising_ASFL}

To approximate a SC material with antiferromagnetic-like nearest-neighbor interactions and
ferromagnetic-like elastic long-range interactions, we here employ the model introduced by S. Miyashita and first used in Refs.~\cite{BROWN201420,PhysRevB.93.064109}. This is a
$L \times L$ square-lattice nearest-neighbor
Ising antiferromagnet with ferromagnetic equivalent-neighbor (aka Husimi-Temperley) interactions.
It is defined by the Hamiltonian,
\begin{equation}\label{def_Hamiltonian_2D_Ising-ASFL}
\mathcal{H} = J \sum_{\langle i,j \rangle}s_{i}s_{j}-HM -\frac{A}{2L^{2}}M^{2} ~,
\end{equation}
with $J>0$.
We name it the two dimensional Ising Antiferromagnetic Short-range and Ferromagnetic Long-range
(2D Ising-ASFL) model.
The first two terms constitute the Wajnflasz-Pick Ising-like model \cite{WAJN71},
in which the pseudo-spin variable $s_i$ denotes the two spin states at site $i$ ($-1$ for LS and $+1$ for HS),
and $M=\sum_{i}s_{i}$ is the pseudomagnetization. The effective field term,
\begin{equation}
H =  \frac{1}{2} (k_{\rm B} T \ln r - D)\;,
\label{eq:heff}
\end{equation}
contains $D > 0$, which is the energy difference between the HS and LS states, and
$r$, which is the ratio between the HS and LS degeneracies. $T$ is the absolute
temperature, and $k_{\rm B}$ is Boltzmann's constant.
(Changing the temperature in the physical SC system therefore corresponds to a combined
change in temperature and effective field in this pseudospin model.
See Figs.~5(a) and 8 of Ref.~\cite{PhysRevB.93.064109}.)

The last term in Eq.~(\ref{def_Hamiltonian_2D_Ising-ASFL}) approximates the elastic long-range
interactions of the SC material
as in Refs.~\cite{PhysRevE.81.011135,PhysRevB.84.054433,PhysRevB.93.064109}.
Since it lowers the energy of more uniform spin-state configurations
(mostly +1 or mostly $-1$) in a quadratic fashion, it is a ferromagnetic term.
Throughout the paper, temperature ($T$), energy ($E$), magnetic field ($H$), and long-range interaction
strength ($A$), will be expressed in
dimensionless units ($|J|=k_{B}=1$).

The order parameters of this model are magnetization ($M$) and staggered magnetization
($M_{s}$). They can be normalized as $m=M/L^{2}$ and $m_{s}=M_{s}/L^{2}$.
If we break the two-dimensional square lattice into two sublattices (A and B),
like the black and white squares
on a chessboard, $m$ and $m_{s}$ can be expressed in terms of the normalized magnetizations
($m_{A}$, $m_{B}$) of these two sublattices as
\begin{eqnarray}
 m &=& (m_{A}+m_{B})/2     \label{def_mag_us} \\
 m_{s}  &=& (m_{A}-m_{B})/2  ~.  \label{def_stgmag_us}
\end{eqnarray}
The usual order parameter for SC materials is the proportion of HS molecules, $n_{\rm HS}$,
which is related to the pseudospin variables as  $n_{\rm HS} = \left( m + 1 \right)/2$.

The equilibrium (stable) and metastable phases at zero temperature were obtained
from the Hamiltonian by simple
ground-state calculations in \cite{PhysRevB.93.064109}. We briefly repeat the results here for
convenient reference, also introducing the following short-hand notation for the low-temperature
ordered phases:

antiferromagnetic (which is doubly degenerate), is called AFM;

ferromagnetic with majority of $s_i = +1$, is called FM$+$;

and ferromagnetic with majority of $s_i = -1$, is called FM$-$.

\noindent
\underline{$A < 8$:}
AFM is stable for $-4 + A/2 < H < 4 - A/2$, metastable against transition to FM+ for
$4 - A/2 < H < 4$, and  metastable against transition to FM$-$ for
$-4 < H < -4 + A/2$.
FM+ is stable for $H > 4 - A/2$, and metastable for transition to AFM or FM$-$ for
$4 - A < H < 4 - A/2$.
FM$-$ is stable for $H < -4 + A/2$, and metastable for transition to AFM or FM$+$ for
$-4 + A/2 < H < -4 + A$.

\noindent
\underline{$A > 8$:}
AFM is never the stable ground state, but it is metastable for $-4 < H < 4$.
FM+ is stable for $H > 0$ and metastable for $4 - A < H < 0$.
FM$-$ is stable for $H < 0$ and metastable for $0 < H < -4 + A$.

\section{Method}\label{sec:method}
\subsection{Obtaining joint density of states}\label{sec:short_DOS}

The results  presented in this paper are all based on the joint DOS, $g(E,M,M_{s})$, determined once for $H=A=0$, which corresponds to a simple square-lattice Ising antiferromagnet.
Using this, the joint DOS for any arbitrary value of $(H,A)$ can be obtained by
\begin{equation}
\label{def_shift_E_DOS}
g(E(H,A),M,M_{s}) = g(E(0,0),M,M_{s})
\end{equation}
where
\begin{equation}
E(H,A) = E(0,0)-HM- \frac{AM^{2}}{2L^{2}}  \ .
\end{equation}
Note that this is an alternative, but equivalent way to express the content of Eq.~(10) in Ref.~\cite{PhysRevE.95.053302}.
This result is based on the fact that all the microstates are equally shifted in energy
when a field-like parameter couples to a function of the global property $M$, as shown in  Eq.~(\ref{def_Hamiltonian_2D_Ising-ASFL}).
With the joint DOS, all thermodynamic quantities can be calculated, as
demonstrated in \cite{PhysRevE.95.053302}. From $g(E,M,M_{s})$ at different $(H,A)$, we can obtain $g(E,M)$ and $g(E,M_{s})$, as shown in Ref.~\cite{CHAN17conference}.

To obtain an accurate  $g(E,M,M_{s})$ at $H=A=0$, the macroscopically constrained
WL method is used \cite{PhysRevE.95.053302,PhysRevE.94.042125}.
With the help of simple combinatorial calculations in the $(M,M_{s})$ space,
the method converts what would otherwise be a time-consuming multi-dimensional
random walk in the $(E,M,M_{s})$ space into many independent, one-dimensional random
walks in $E$, each constrained to a fixed value of $(M,M_{s})$. Through further,
symmetry based simplifications \cite{PhysRevE.95.053302}, the method can obtain an accurate
estimate of $g(E,M,M_{s})$ in a relatively short time.

As the details of how to arrive at these results have already been presented in \cite{PhysRevE.95.053302},
here we simply focus on the physics of the model SC material as $A$ is changed.
All the phase diagrams, free-energy landscapes, and probability densities shown in
Sec.~\ref{sec:phase_diagram} are  obtained with $L=32$.

\subsection{From joint density of states to thermodynamic quantities}\label{sec:maths_phase_diagram}

We define the constrained partition function of any macrostate $(m,m_{s})$ as
\begin{equation}\label{def:partition function_pt}
Z_{m,m_{s}}=\sum_{E} g(E,m,m_{s}) e^{-E/T} ~.
\end{equation}
The overall partition function of the system is then
\begin{equation}\label{def:partition function_all}
Z_{\rm{all}}=\sum_{m,m_{s}} Z_{m,m_{s}} ~.
\end{equation}
The joint probability of finding the system in a macrostate $(m,m_{s})$ is
\begin{equation}\label{def:joint_Prob_density}
P(m,m_{s})\Delta m \Delta m_{s}=\frac{Z_{m,m_{s}}}{Z_{\rm{all}}} ~,
\end{equation}
where $\Delta m$, $\Delta m_{s}$ are the order-parameter step sizes,
both chosen to be the same value, around $0.03$.
The free energy of macrostate $(m,m_{s})$ is
\begin{equation}\label{def:Free_energy_pt}
F(m,m_{s})= -T \ln Z_{m,m_{s}} ~.
\end{equation}
We will plot these quantities in terms of $(m_{A},m_{B})$ which have a one-to-one relation
with $(m,m_{s})$  (see Eqs.~(\ref{def_mag_us}) and (\ref{def_stgmag_us})).

Summing over the contributions of the joint probability (Eq.~(\ref{def:joint_Prob_density})) in one direction,
we obtain the marginal probability densities as
\begin{eqnarray}
 P(m)\Delta m &=& \frac{\sum_{m_{s}} Z_{m,m_{s}}}{Z_{\rm{all}}}      \label{def:marginal_prob_m} \\
 P(m_{s})\Delta m_{s}  &=& \frac{\sum_{m}  Z_{m,m_{s}}}{Z_{\rm{all}}}  ~.  \label{def:marginal_prob_ms}
\end{eqnarray}
With these densities,
we can calculate the expectation values of the order parameters and other quantities.
We can express the free energy in terms of one order parameter as
\begin{eqnarray}
 F(m) &=& -T \ln \sum_{m_{s}} Z_{m,m_{s}}     \label{def:marginal_F_m} \\
 F(m_{s})  &=& -T \ln \sum_{m}  Z_{m,m_{s}} ~.  \label{def:marginal_F_ms}
\end{eqnarray}

The presence of the long-range interaction induces metastable phase regions in the phase diagrams.
 A very important point is that
 when we consider values of $(T,H,A)$ lying in those regions, the stable phase will be the phase
that has larger total area in the marginal probability density, rather than the phase that shows
the higher peak. Systems lying on the coexistence line between two phases will have equal areas
in the marginal probability density.

In a free-energy contour plot or joint probability density plot, against $m$ and $m_{s}$ (or against $m_{A}$ and $m_{B}$), the
region around $(m,m_{s})=(1,0)$ [or $(m_{A},m_{B})=(1,1)$]  corresponds to the FM+ phase.
Similarly, the region around $(m,m_{s})=(-1,0)$ [or $(m_{A},m_{B})=(-1,-1)$] corresponds to the FM$-$ phase.
The region around $(m,m_{s})=(0,1)$ [or $(m_{A},m_{B})=(1,-1)$] corresponds to the AFM+ phase,
and the region around $(m,m_{s})=(0,-1)$ [or $(m_{A},m_{B})=(-1,1)$] corresponds to the AFM$-$ phase.
Finally, the region around $(m,m_{s})=(0,0)$ [or $(m_{A},m_{B})=(0,0)$] corresponds to the disordered phase.
However, these are just the most extreme cases. Some AFM phases have significant ferromagnetic properties, and some FM phases may be quite disordered.

In our model, for a particular  $(T,H,A)$ triple, if the system can exist as a disordered phase, it cannot exist as an AFM phase, and vice versa. However it may happen that
a disordered phase shows strong AFM properties.
Changing $(T,H,A)$ may let the system change from one phase to another through a continuous phase transition, as it crosses the critical line between the two phases. In the Ising-ASFL model, a critical line only exists between the disordered phase and the AFM phase. The phase boundary between the ferromagnetic phase and the disordered phase is a coexistence line, and it ends with a mean-field critical point for sufficiently strong long-range interaction $A$. This critical point is located where the two spinodal lines meet.

The expectation values of the two order parameters can be obtained easily as
\begin{eqnarray}
 \langle m \rangle &=&   \sum_{m} m P(m) \Delta m   \label{def:exp_m} \\
 \langle m_{s} \rangle  &=&  \sum_{m_{s}} m_{s} P(m_{s}) \Delta m_{s}  ~.  \label{def:exp_ms}
\end{eqnarray}
As the two AFM phases always exist in pairs and the probability of finding the system in both
are the same, $\langle m_{s} \rangle=0$.

\begin{figure}
\includegraphics[width=0.5\textwidth]{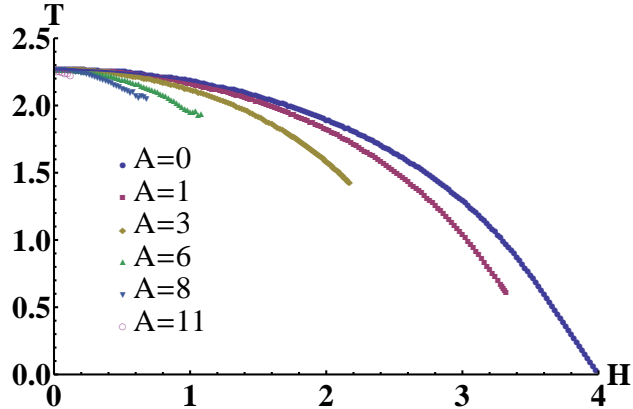}
\caption{Critical lines for six different values of $A$. The critical lines are obtained by increasing $H$ in steps of  0.01 or 0.02, then performing a temperature scan, choosing $\Delta T$ to be 0.001 to 0.005, and locating the critical line by choosing the point that gives the cumulant value (refer to Eq.~(\ref{def_cumulant_ms})) closest to 0.61 \cite{0305-4470-26-2-009,PhysRevE.70.056136,0305-4470-38-44-L03,Salas2000}. If, at a certain $H$, all the cumulant values obtained for different $T$  deviate from $0.61$ by more than $0.01$, the critical line is considered to have terminated. When calculating the cumulants, all the phase points that belong to the metastable FM+ or FM$-$ phase were disregarded, i.e., the critical line separates the AFM phases from the
disordered phase. The analytically approximated critical line for the antiferromagnetic Ising model, $A=0$, from \cite{WU1990123} is also plotted. Within  the resolution of this figure, it coincides with our data points for $A=0$.
Adding a ferromagnetic long-range interaction $A>0$ favors the appearance of the ferromagnetic phase, and thus pushes the critical line towards lower values of $|H|$. Moreover, the critical line also terminates at higher $T$ as $A$ increases.
The critical lines are symmetric about $H=0$.
}\label{ASFL_phase_diagram_criticalline}
\end{figure}

As $\langle m_{s} \rangle=0$, we define the corresponding fourth-order Binder cumulant as \cite{Landau_simulation_book,PhysRevLett.47.693, PhysRevB.30.1477,PhysRevB.34.1841},
 \begin{equation}\label{def_cumulant_ms}
u_{m_{s}}=1-\frac{ \langle m_{s}^{4} \rangle }{3 \langle m_{s}^{2} \rangle^{2} } ~.
\end{equation}
Here we only define the cumulant for the order parameter $m_{s}$, as only the critical line will be located by the cumulant. When we take the ensemble average, we have to exclude all the phase points that belong to the metastable FM+ or FM$-$ phase. That is, when we look at $F(m)$, if we find more than one minimum (i.e. more than one phase are found), we neglect the states that have values of $|m|$ greater than the separating value of $m$.
The critical line in this model is commonly accepted to be in the Ising universality class \cite{1742-5468-2016-3-033107}, which (assuming isotropy, periodic boundary conditions,
and a square shape as in the present study)
has a cumulant fixed-point value of $0.6106924(16)$ \cite{0305-4470-26-2-009,PhysRevE.70.056136,0305-4470-38-44-L03,Salas2000}.
We therefore locate the critical line by finding the phase point within a temperature range where the cumulant is close to $0.61$, and does not deviate from $0.61$ by more than $0.01$.
The resulting critical line for $A=0$ is included in Fig.\ \ref{ASFL_phase_diagram_criticalline} together with
the analytically approximated critical line for the pure square-lattice
Ising antiferromagnet in the thermodynamic limit from Ref.\ \cite{WU1990123}.
Within  the resolution of this figure, our $L=32$ data  coincide with
this highly accurate approximation.

The variance of the order-parameter $m$, which is proportional to the susceptibility times the temperature,
\begin{equation}\label{def_susceptibility0}
\textrm{var}(m)=\chi_{m}T = L^{2}( \langle m^{2}\rangle - \langle m \rangle^{2} )  ~,
\end{equation}
is considered as we use its maxima to separate the FM$\pm$ phases from the disordered and AFM phases. All the coexistence lines that we show are located by using this quantity.
Note that this quantity is very difficult to measure through importance-sampling MC,
while our approach can directly calculate it using $g(E,M,M_{s})$.
Further details on the method are given in Ref.~\cite{PhysRevE.95.053302}.

In next section, we consider the phase diagrams for different values of $A$
and study selected phase points. These are the main results of the present paper. Notice that all the phase diagrams are symmetric about the $T$ axis, with an exchange between FM$+$ and FM$-$. For $A=0$, the model reduces to the standard square-lattice antiferromagnetic Ising model \cite{1742-5468-2016-3-033107,PhysRevE.95.053302}.

\section{Phase diagrams}\label{sec:phase_diagram}
\subsection{Weak long-range interaction, $A=1,4$} \label{sec:small_A}

\begin{figure}
\includegraphics[width=0.8\textwidth]{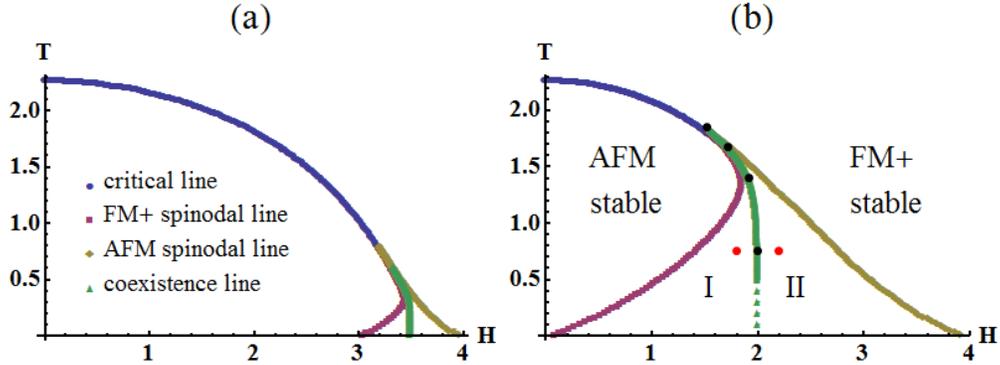}
\caption{Phase diagrams for cases of weak long-range interaction $A$, with (a) $A=1$ and (b) $A=4$. The meaning of each line is shown in the legend in (a), and each phase region is labeled in (b). Region I represents stable AFM phase with metastable FM$+$ phase, and region II represents stable FM$+$ phase with metastable AFM phase.
The metastable regions grow as $A$ increases. Notice that when $T$ is small, the coexistence lines are straight lines at constant $H$.
The dots mark phase points studied in the next few figures.
Error bars are everywhere smaller than or comparable to the symbol size.
Unless otherwise noted, this is also the case for all other phase diagrams shown in this paper.
}\label{ASFL_phase_diagram_A1and3and4}
\end{figure}

\begin{figure}
\includegraphics[width=0.5\textwidth]{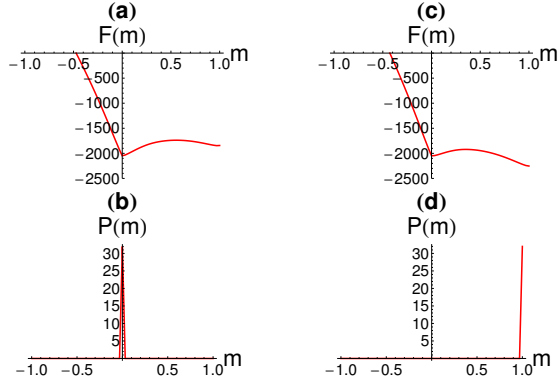}
\caption{Free energy, $F(m)$ (Eq.~(\ref{def:marginal_F_m})), and marginal probability density, $P(m)$ (Eq.~(\ref{def:marginal_prob_m})), for two points lying in two different metastable regions with $A=4$ in Fig.~\ref{ASFL_phase_diagram_A1and3and4}(b). Both points have the same temperature, $T=0.75$, and are equidistant from the coexistence line. (a)-(b) show $H=1.8$
(region I in Fig.~\ref{ASFL_phase_diagram_A1and3and4}(b)), which has a stable AFM phase and a metastable FM$+$ phase. (c)-(d) show $H=2.2$
(region II in Fig.~\ref{ASFL_phase_diagram_A1and3and4}(b)), which has a stable  FM$+$ phase and a metastable AFM phase. Note that in both cases, the metastable phases are easily observed in the free energy, but their corresponding peaks in the marginal probability density are too small compared to the peaks of the stable phases, such that they are not observed in (b) and (d). Both phase points would be located in the AFM phase region for $A=0$. Adding the long-range interactions creates a local free-energy
minimum in the FM$+$ region, and thus brings out the metastable phase.
}\label{ASFL_A4_2metastable_free_E_and_ProbD}
\end{figure}

\begin{figure}
\includegraphics[width=1\textwidth]{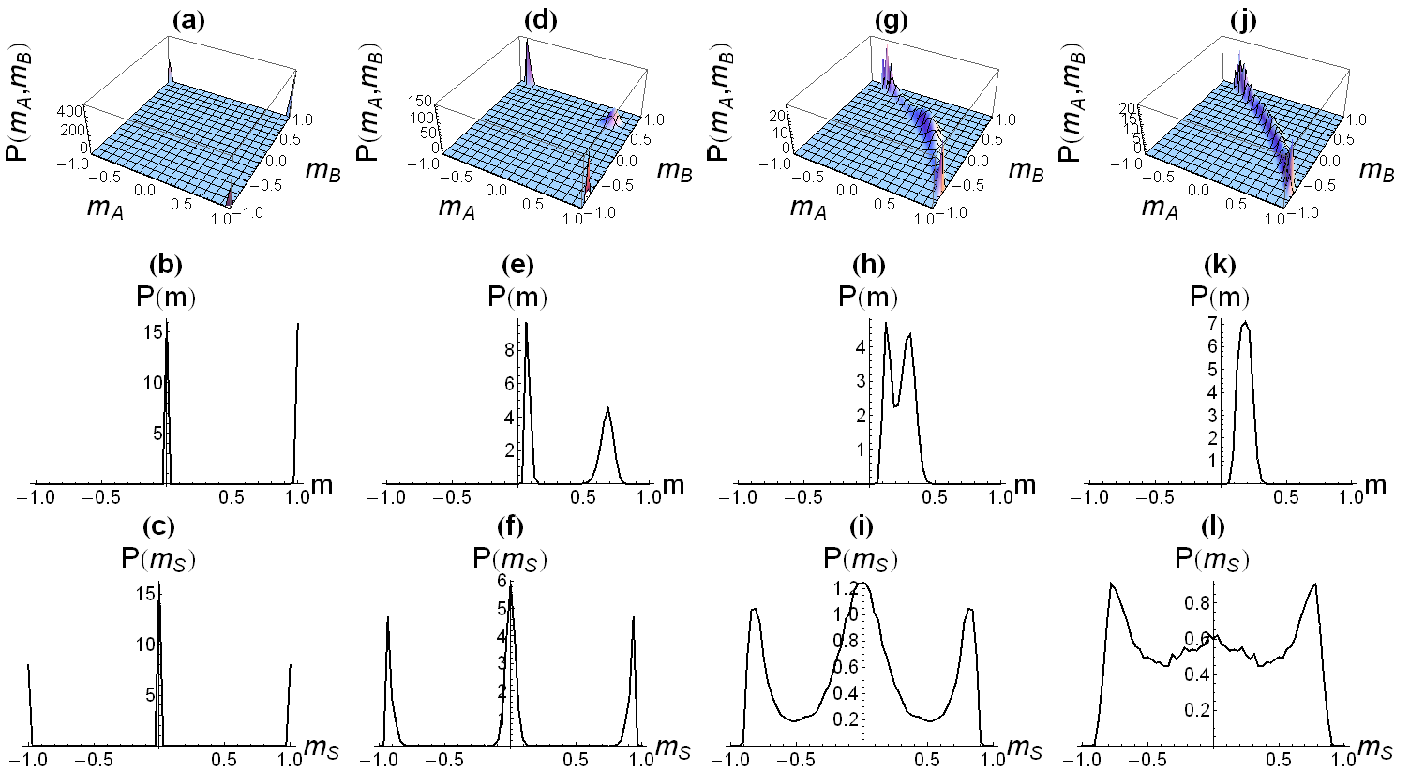}
\caption{
Changes of the probability densities, $P(m_{A},m_{B})$ (Eq.~(\ref{def:joint_Prob_density})), and marginal probability densities, $P(m)$ (Eq.~(\ref{def:marginal_prob_m})) and $P(m_{s})$ (Eq.~(\ref{def:marginal_prob_ms})), when moving along the coexistence line starting from a low $T$ toward the tricritical point for $A=4$ as marked in
Fig.~\ref{ASFL_phase_diagram_A1and3and4}(b).
The surface plots and the graphs,
in (a)-(c) show $(H,T)=(2.0005,0.75)$, which has the same temperature as in Fig.~\ref{ASFL_A4_2metastable_free_E_and_ProbD}, in (d)-(f) show $(H,T)=(1.916,1.4)$, in (g)-(i) show $(H,T)=(1.716,1.68)$, and in (j)-(l) show $(H,T)=(1.5147, 1.85)$ which is the tricritical point.
(a)-(c) show the typical pattern when $T$ is low. The system is found in $m=+1$ (FM$+$ phase) or $|m_{s}|=1$ (AFM phase), so that sharp peaks are found at these points. The system is equally probable in the FM$+$ phase and the AFM phase, i.e. the areas under the two peaks in (b) are the same, and the sum of the areas of the two peaks in (c) at $|m_{s}|=1$ is equal to the area of the peak at  $|m_{s}|=0$. Further increase in $T$ along the coexistence line makes the AFM peaks and the FM$+$ peak move toward each other, and the peaks become wider, as shown
in (d)-(f). Even further increase in $T$ makes the peaks coalesce  as shown in (g)-(i). As two peaks are still observed in (h), we regard them as two different phases, between which this small system can fluctuate easily. Notice in (g) that the FM$+$ phase has spread out significantly. Finally the three peaks join together in (j)-(l), and we regard this point as the tricritical point.
}\label{ASFL_A4coexist}
\end{figure}

It is reasonable to assume that adding a ferromagnetic long-range interaction $A$ to the pure antiferromagnet must favor the appearance of the ferromagnetic phases, and thus push the critical line towards lower values of $|H|$. Figure \ref{ASFL_phase_diagram_criticalline} supports this assumption. Moreover, the critical lines also terminate at lower $|H|$ and higher $T$ for larger $A$. The phase diagrams in Fig.~\ref{ASFL_phase_diagram_A1and3and4} show that the critical lines
end with the appearance of a metastable region in the phase diagram, and that the metastable region grows as $A$ increases.
All phase diagrams shown in this paper are symmetric under simultaneous reversal of $H$ and $m$.
Error bars including statistical and finite-size errors are included with every data point in this and
all subsequent phase diagrams. With the exception of Fig.\ \ref{ASFL_A7horn},
they are everywhere smaller than the symbol size.
A discussion of how the errors were estimated is found in Appendix \ref{sec:finite_size_error}.

Introducing the long-range interaction $A$ with the $M^{2}$ term makes it much weaker than the $HM$ term for small $M$, so that the long-range interaction effect is negligible
when $H$ and $A$ are small, and so it does not
significantly affect the critical temperature near $H=0$. On the other hand,
when we increase $H$, the $M^{2}$ term will eventually be larger than the $M$ term, and finally causes a local free-energy minimum to show up in the FM$+$ region, corresponding to a metastable FM$+$ phase region in the phase diagram (Figs.~\ref{ASFL_A4_2metastable_free_E_and_ProbD} (a) and (b)). A new FM$+$ peak also appears in the  joint probability density ($P(m_{A},m_{B})$) and marginal probability densities ($P(m)$ and $P(m_{s})$). One peak may be much smaller
than the other, such that it may not be easy to discover the presence of metastability through looking at the probability density (Figs.~\ref{ASFL_A4_2metastable_free_E_and_ProbD} (b) and (d)). Notice that although one phase may have much smaller probability density than the other, the lifetimes
for these metastable phases  increase exponentially with system volume, $e^{cL^2}$ for a
two-dimensional system, so that they are still macroscopic, and thus cannot be neglected \cite{PhysRevE.81.011135,arXiv:cond-mat/9407027,:/content/aip/journal/jcp/145/21/10.1063/1.4959235}.

The AFM and FM+ phases are separated by the coexistence line in the metastable region, and we observe that when $T$ is low, the coexistence line is a practically straight line at constant $H$ in the phase diagram.
Note that this result is different from Rikvold \textit{et al.}'s former result \cite{PhysRevB.93.064109} for $A=4$, which indicates a reentrant behavior of the coexistence line at low $T$.
This discrepancy is probably due to incomplete ergodicity in the importance-sampling
MC with mixed initial conditions used in Ref.\ \cite{PhysRevB.93.064109}.

For any point lying on that straight vertical segment of the coexistence line, as in Fig.~\ref{ASFL_A4coexist} (a)-(c), the coexisting AFM phases and the FM+ phase are located at their extreme locations, i.e., $m=+1, m_{s}=\pm1$.
 Increasing $T$ bends the coexistence line toward lower $|H|$ values. Simultaneously, the AFM phases and the FM$+$ phase move away from the extreme positions and towards each other, as shown in Fig.~\ref{ASFL_A4coexist} (d)-(f). The coexistence line finally joins the critical line at the tricritical point, where the two AFM phases and the FM+ phase become indistinguishable at the continuous phase transition point. Figure ~\ref{ASFL_A4coexist} (g)-(i) represent a point lying on the coexistence line, below the tricritical point. We see from the joint probability density in Fig.~\ref{ASFL_A4coexist} (g) that the ferromagnetic phase and the AFM phases are coalescing. However, the marginal probability along the $m$ axis in Fig.~\ref{ASFL_A4coexist} (h) still has two peaks. We therefore regard the system as in AFM/FM$+$ coexistence, with this small system fluctuating easily between the two phases.
Extrapolation of the end points of the two spinodal lines gives the merging temperature, which corresponds to the tricritical point.
 When the two spinodal lines merge, the distance between them ($\Delta H$) varies against temperature as \cite{Newman1980}
\begin{equation}\label{spinodal_vs_temp}
(\Delta H)^{2/3} \propto T_{x}-T ~,
\end{equation}
where $T_{x}$ represents the tricritical or critical temperature, where the coexistence line ends.
After obtaining the tricritical temperature, we can estimate the tricritical field as the average of the extrapolation points of the two spinodal lines.
 Figure~\ref{ASFL_A4coexist} (j)-(l) show data at the tricritical point for $A=4$, where the AFM phases and the FM$+$ peak finally join together into one phase.

\subsection{Medium long-range interaction, $A=6,7,8$} \label{sec:medium_A}

\begin{figure}
\includegraphics[width=1\textwidth]{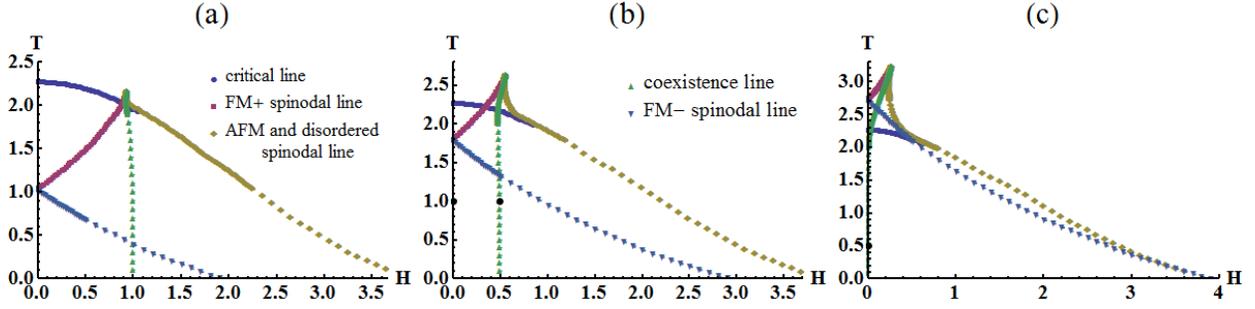}
\caption{Phase diagrams for $A=6$, $7$, and $8$. The FM$-$ metastable phase, which only exists
for $H<0$ for small $A$, now also exists in the $H>0$ region, so a FM$-$ spinodal line shows up in these graphs. This FM$-$ spinodal line moves toward larger $T$ and $H$ when $A$ increases, and the metastable  FM$-$  region grows.
  The coexistence line crosses the critical line at the critical end-point, and finally meets two spinodal lines at a new mean-field critical point, which brings out a new metastable region, the horn region. A closer look at the horn regions for $A=7$ and $8$ are shown in Figs.~\ref{ASFL_A7horn} and \ref{ASFL_A89horn}(a), respectively.
When $A$ increases, the mean-field critical temperature rises, which makes the area of the horn region increase. The coexistence line moves towards the $T$-axis as $A$ increases, which makes the stable AFM region shrink and the stable FM$+$ region grow.
The black  dots mark phase points studied below.
}\label{ASFL_phase_diagram_A6and7and8}
\end{figure}

\begin{figure}
\includegraphics[width=0.5\textwidth]{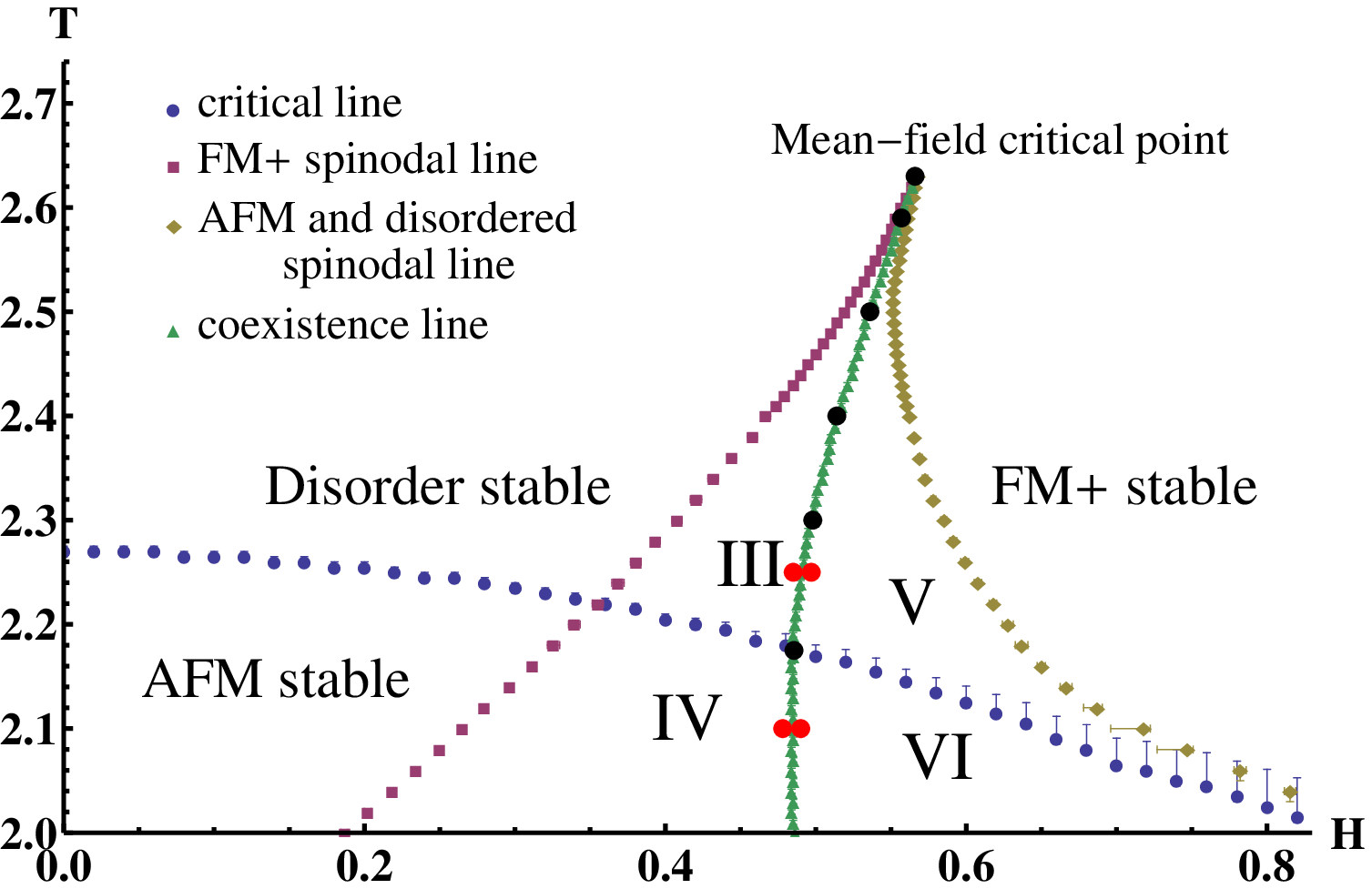}
\caption{The horn region of the phase diagram for $A=7$.
Region III is disordered phase stable with metastable FM$+$ phase, region IV is AFM phase stable with metastable FM$+$,  region V is FM$+$ phase stable with metastable disordered phase, and region VI is FM$+$ phase stable with metastable AFM phase. Observe that the coexistence line turns toward stronger $|H|$ when approaching the mean-field critical point. This is because higher temperature favors the disordered phase. The black and red dots mark phase points studied in the next few figures.
At this enlarged scale, error bars are visible in the lower right part of the figure.
}\label{ASFL_A7horn}
\end{figure}

As mentioned above for small $A$, moving along the coexistence line toward the critical line,
one approaches a tricritical point, where the two AFM phases and the FM phase
become indistinguishable. Below the tricritical temperature, the three phases are distinct.
Then it is reasonable to expect that, if $A$ is big enough, the two AFM phases may combine
into one disordered phase at a lower $T$ than the one where they further combine
with the FM phase.
In this scenario, we will find that the critical line, which represents the AFM/disordered
phase transition, intersects the coexistence line at a critical end-point, and new
metastable regions (horn regions) emerge in the phase diagram as shown for
$A=6,7$, and $8$ in Fig.~\ref{ASFL_phase_diagram_A6and7and8}.

Figurte \ref{ASFL_A7horn} is a closer look at the horn region for $A=7$.
The coexistence line separates the FM phase from the AFM phases at low $T$. After passing through the critical end-point, it separates the FM phase from the disordered phase. At a higher $T$, it ends in a mean-field critical point, where the disordered and FM phases become indistinguishable.

\begin{figure}
\includegraphics[width=0.5\textwidth]{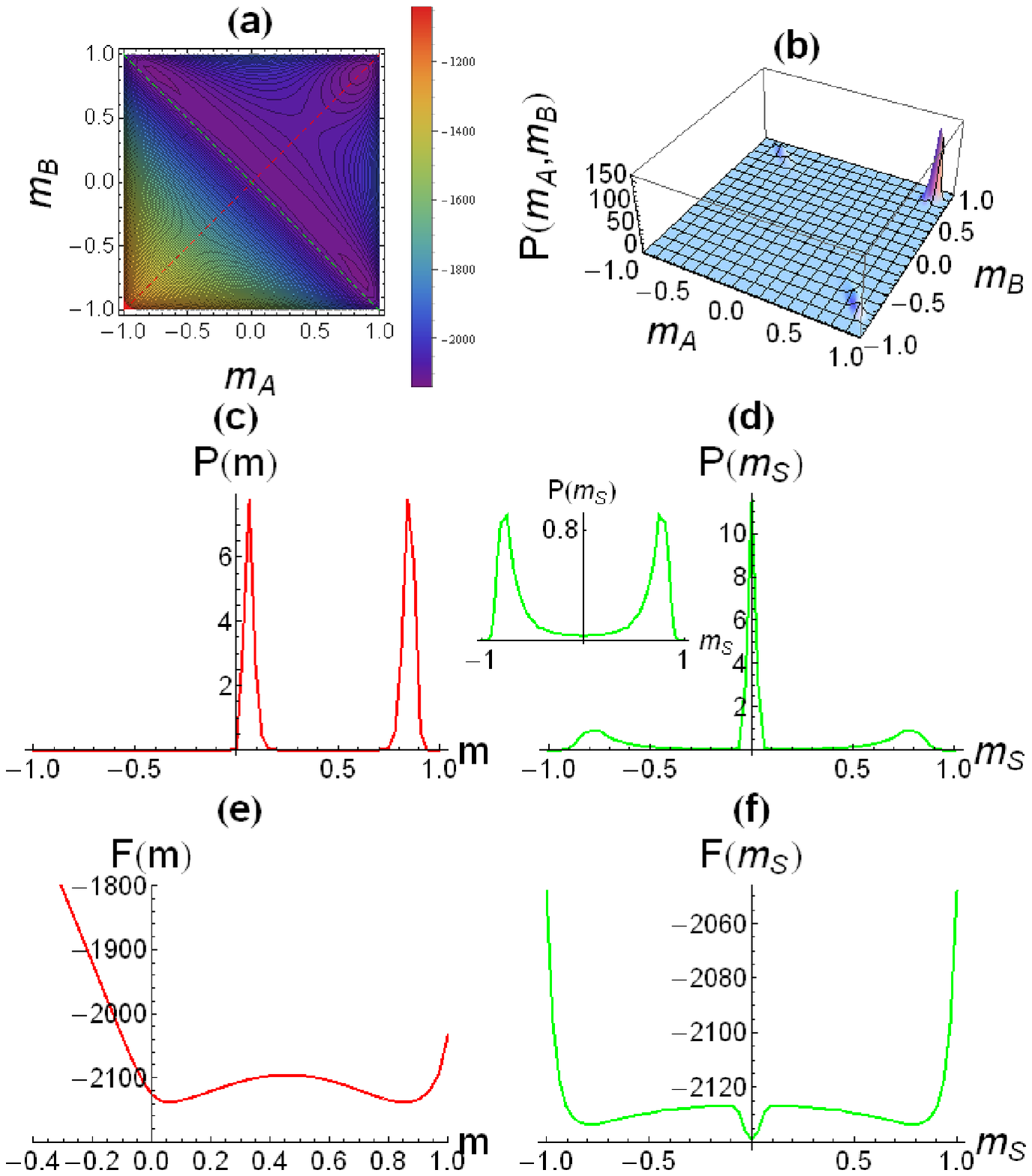}
\caption{Data for $A=7$, $H=0.4855$, $T=2.175$, which is approximately
the critical end-point in the phase diagram of Fig.~\ref{ASFL_A7horn}. (a) shows the free-energy contour, with the red dotted diagonal representing the $m$ axis, and the green dotted diagonal representing the $m_{s}$ axis. (b) shows the corresponding joint probability density
as in Eq.~(\ref{def:joint_Prob_density}). (c) and (d) show the marginal probabilities expressed in Eqs.~(\ref{def:marginal_prob_m}) and (\ref{def:marginal_prob_ms}). (e) and (f) show the free energies in Eqs.~(\ref{def:marginal_F_m}) and (\ref{def:marginal_F_ms}). The inset in (d) shows $P(m_{s})$ after removing the effect from the FM$+$ phase.  A critical end-point is the intersection of the critical line and the coexistence line, and has properties of both lines.   Since it is on the critical line, (b) and the inset in (d)
 show that the AFM peaks are connected through the middle disordered region as it corresponds to a continuous phase transition between the AFM phases and the disordered phase. Since it is on the coexistence line, (c) and (e) show that the combined AFM/disordered phase is equally probable as the FM+ phase.
}\label{ASFL_A7criticalend}
\end{figure}

\begin{figure}
\includegraphics[width=0.5\textwidth]{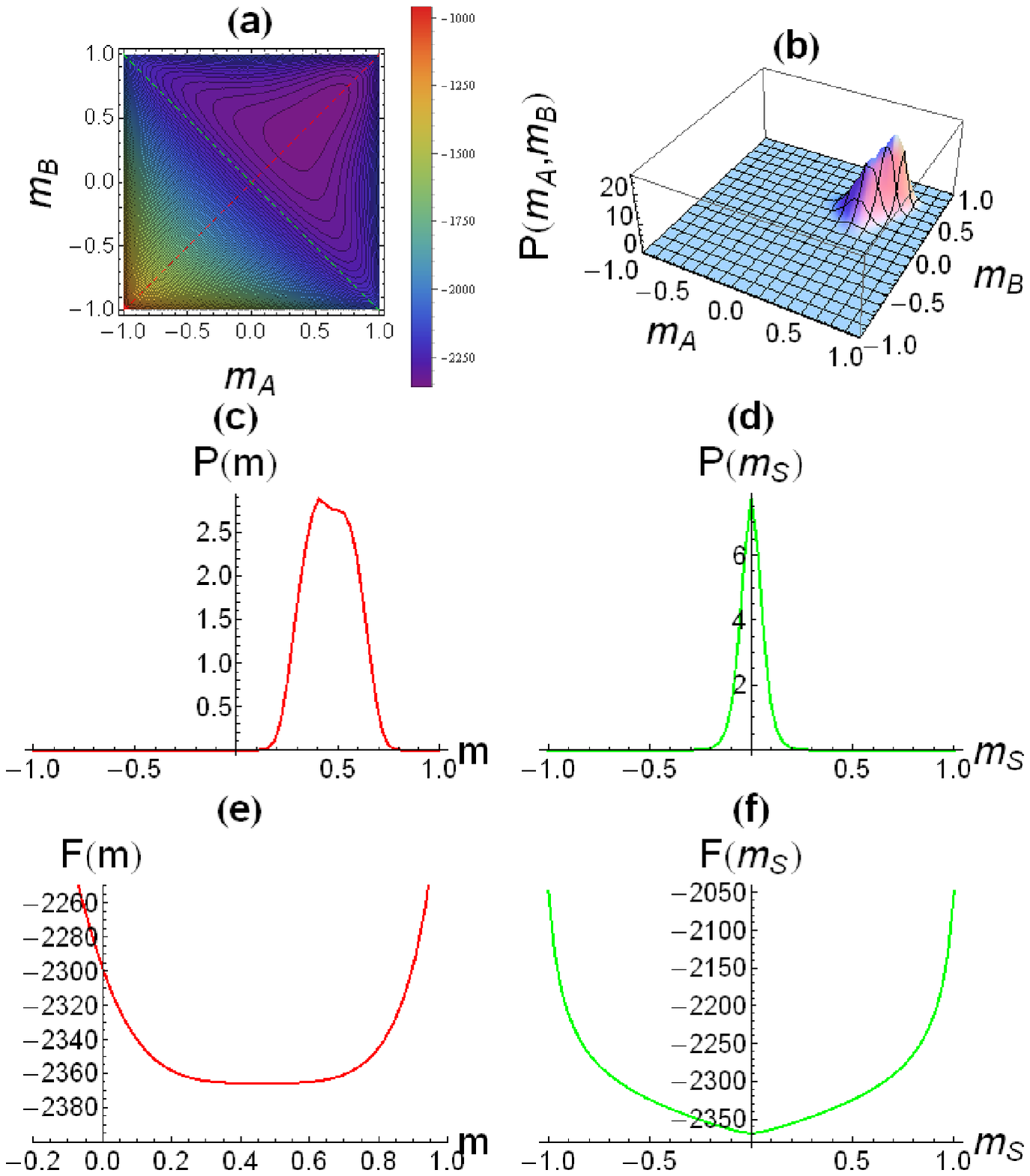}
\caption{$A=7$, $H=0.566$, $T=2.63$. A point very close to the mean-field critical point for $A=7$, where the disordered phase and the FM+ phase have merged together as one phase, as shown in (c).
}\label{ASFL_A7criticalpt}
\end{figure}

Figure \ref{ASFL_A7criticalend} shows the case near the critical end-point. As this point is the intersection of the critical line and the coexistence line, it has properties of both lines. Since it is on the coexistence line, the combined AFM/disordered
 phase is equally probable as the FM$+$ phase, as shown in (c) and (e). Since it is on
the critical line, the AFM peaks are connected through the middle disordered region as it corresponds to a continuous phase transition between the AFM phases and the disordered phase (shown in (b)).
For the marginal probability density function $P(m_{s})$, if we remove the contribution from the FM$+$ phase as shown in the inset, the height ratio between a AFM peak to the central point in the middle between the two peaks is around $26/1$, which is close to the established value of about $22/1$ \cite{0305-4470-21-1-028}.
 Figure \ref{ASFL_A7criticalpt} shows a point close to the mean-field critical point at $(H,T)=(0.566,2.63)$ for $A=7$, where we see that the two peaks in $P(m)$ have coalesced into one single peak. We note that the position of the critical point found here is consistent with the one found in Ref.~\cite{PhysRevB.93.064109} by importance-sampling MC with system sizes up to $L=1024$, $H=0.561(1)$ and $T=2.61(1)$.

\begin{figure}
\includegraphics[width=1\textwidth]{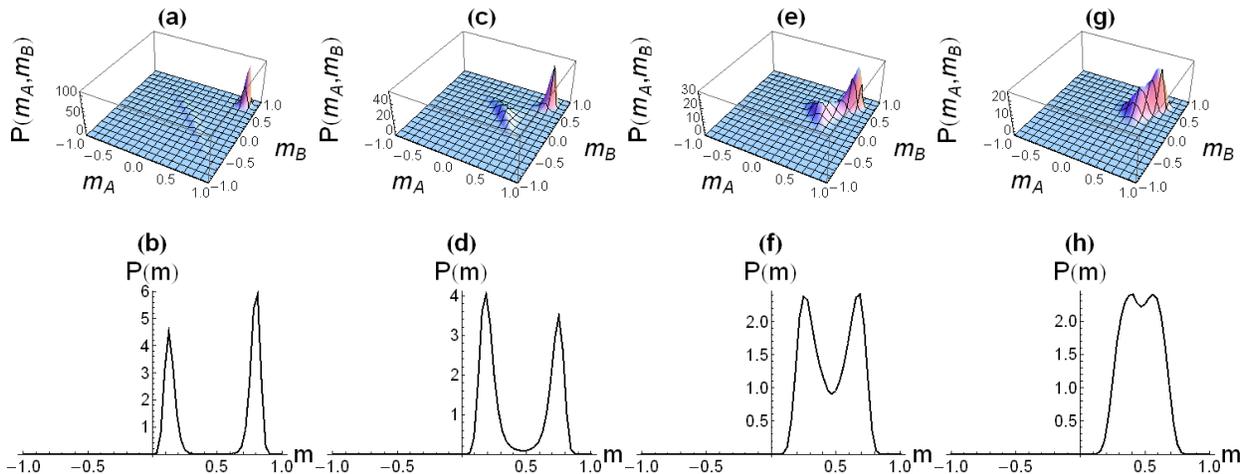}
\caption{Changes of the joint probability densities, $P(m_{A},m_{B})$, and marginal
density, $P(m)$, when moving along the coexistence line starting from a point above the critical
end-point, towards the mean-field critical point for $A=7$ (Fig.~\ref{ASFL_A7horn}).
Parts
(a)-(b) show $(H,T)=(0.498,2.3)$, (c)-(d) show $(H,T)=(0.514,2.4)$,
(e)-(f) show $(H,T)=(0.536,2.5)$, and (g)-(h) show $(H,T)=(0.557,2.59)$,
as marked in  Fig.~\ref{ASFL_A7horn}.
All the graphs for $P(m)$  have two peaks, representing the disordered phase and the FM$+$ phase. Note that $P(m_{A},m_{B})$ may only show one peak as in (g), as long as $P(m)$ has two peaks as in (h), there are still two peaks. Phase points lying on the coexistence line show equal areas below
 the two peaks in $P(m)$, and show maxima in the order-parameter variance (Eq.~(\ref{def_susceptibility0})). From $P(m_{A},m_{B})$ we see that the
disordered phase becomes less dispersed as we move towards the mean-field critical point. Moreover, the disordered  phase and the FM$+$ phase peaks
are moving closer to each other and start coalescing.
}\label{ASFL_A7along_coexistence}
\end{figure}

\begin{figure}
\includegraphics[width=1\textwidth]{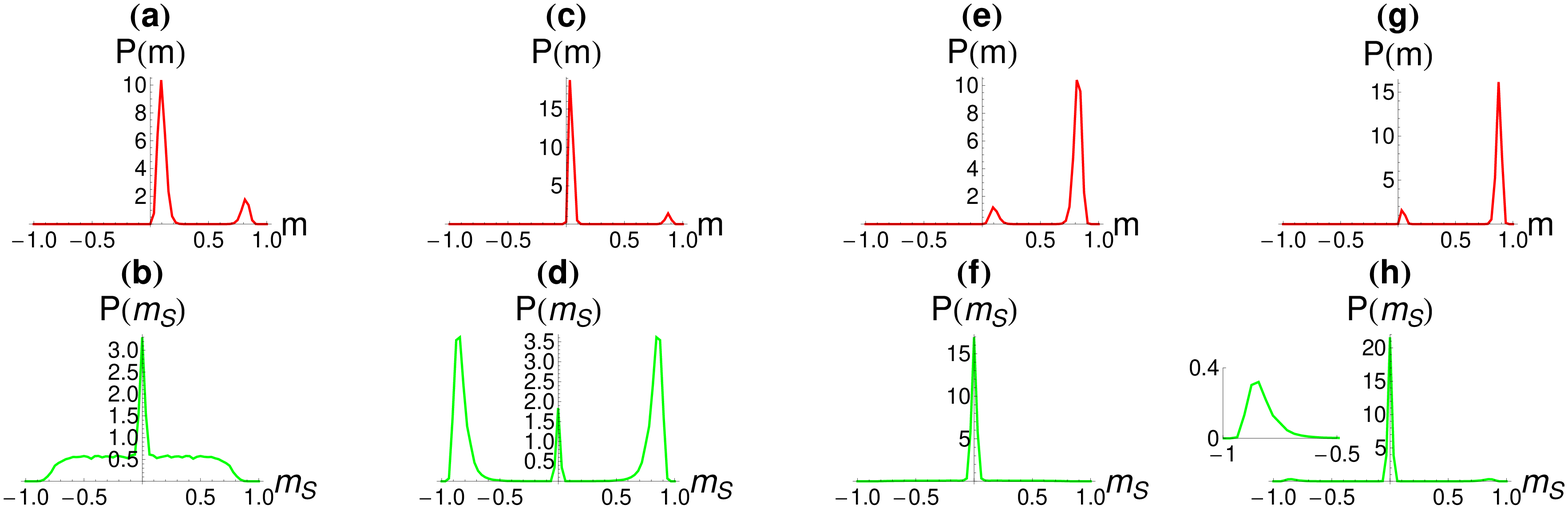}
\caption{Marginal probability densities, $P(m)$ and $P(m_{s})$, for the four red selected phase points lying in the four different metastable regions in Fig.~\ref{ASFL_A7horn} for $A=7$. All the points are equidistant from the coexistence line. (a)-(b) show $(H,T)=(0.485,2.25)$ lying in Region III (stable disordered phase with metastable FM$+$ phase),  (c)-(d) show $(H,T)=(0.478,2.1)$  lying in Region IV (stable AFM phase with metastable FM$+$ phase), (e)-(f) show $(H,T)=(0.497,2.25)$  lying in Region V (stable FM$+$ phase with metastable disordered phase), and (g)-(h) show $(H,T)=(0.49,2.1)$  lying in Region VI (stable FM$+$ phase with metastable AFM). Note that in (h), there are two very small peaks near $m_{s}=\pm 0.85$ corresponding to the AFM phase, and the inset shows one of these.
}\label{ASFL_A7four_pt}
\end{figure}

Figure \ref{ASFL_A7along_coexistence} shows results as we move along the coexistence line to a point near the mean-field critical point. The disordered phase peak gradually contracts to $m_{s}=0$ as the AFM fluctuations weaken (refer to the first row of the figure), and the FM$+$ peak slowly merges with the disordered phase peak until only one peak is left in the marginal probability along the $m$ direction (refer to the second row of the figure).
   We see that the two peaks in the marginal probability density, $P(m)$, along the FM axis, which correspond to two different phases, become less sharp and merge. Note that the joint probability density in (g) seems to show only one peak, but after summing up all the contributions from different $m_{s}$, the marginal probability density in (h) shows two peaks, and we still regard them as two phases even though they are strongly connected by fluctuations.

Figure \ref{ASFL_A7four_pt} shows the results observed at four points that are equidistant from the coexistence line, but lie in four different phase regions, with the critical end-point nearly at the
center, as shown by the four red dots in Fig.~\ref{ASFL_A7horn}. Parts (a)-(b) show a point lying in region III, which has stable disordered phase and metastable FM$+$ phase;  (c)-(d) show a point lying in region IV, which has stable AFM phase and metastable FM$+$ phase;  (e)-(f) show a point lying in region V, which has stable FM$+$ phase and metastable disordered phase; and  (g)-(h) show a point lying in region VI, which has stable FM$+$ phase and metastable AFM phase.

\begin{figure}
\includegraphics[width=0.5\textwidth]{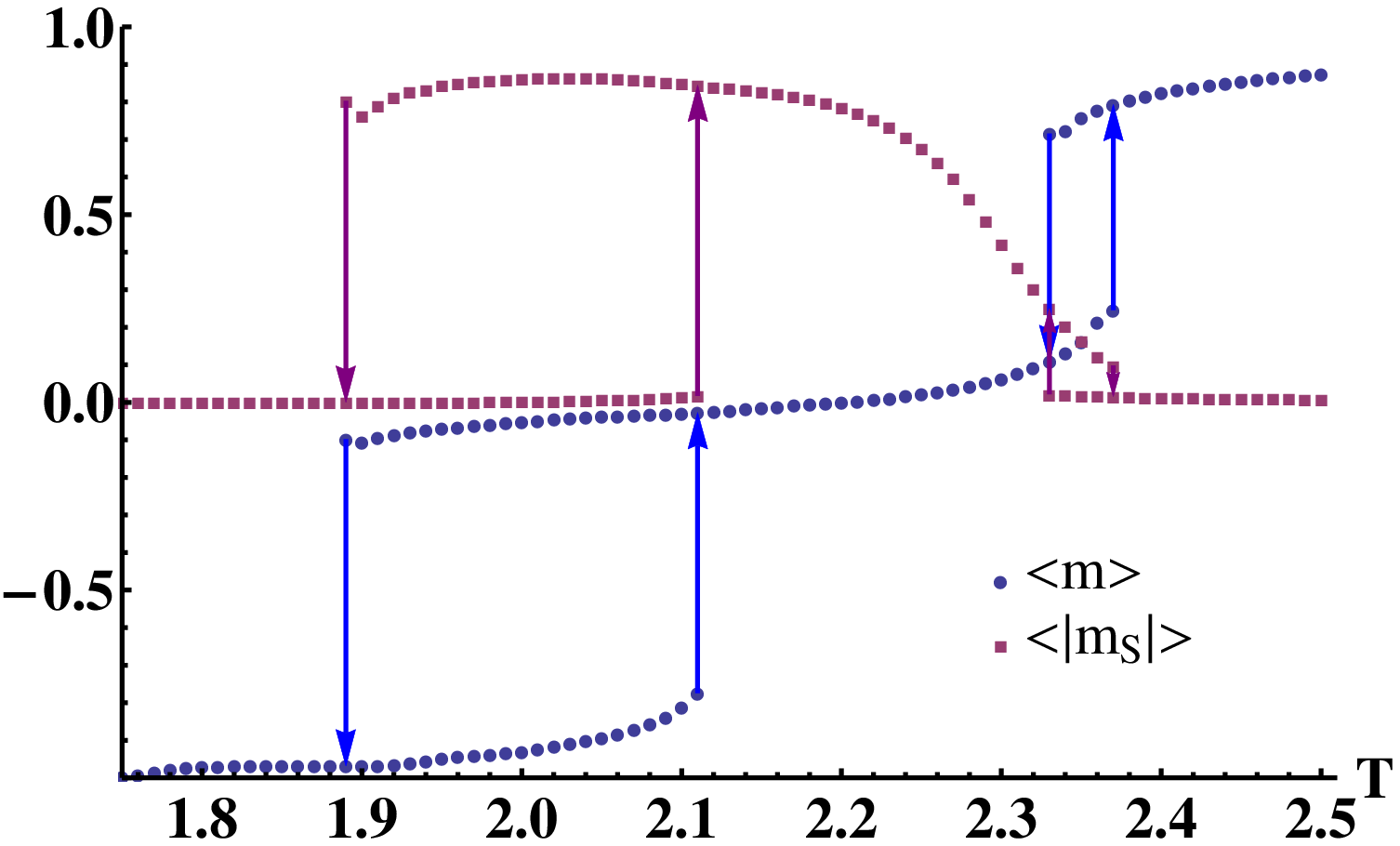}
\caption{
Asymmetric, two-step hysteresis loops for $A=7$, reminiscent of experimental results for several SC
materials \protect\cite{KOPP82,ZELE85,PETR87,BOUS92,JAKO92,REAL92,BOIN94,BOLV96,CHER04,BONN08,PILL12,BURO10,LIN12,KLEI14,CHER03,HUBY04,SHAT15,BROO15}.
The phase point moves back and forth along a path between $(H,T) = (1.0,2.5)$ and
$(-1.5,1.75)$ in Fig.\ \protect\ref{ASFL_A7horn},
corresponding to the parameters $\ln r=20/3$ and $D=44/3$ in Eq.~(\ref{eq:heff}).
The narrow high-temperature loop corresponds to the crossing of the horn region, while the wide
low-temperature loop corresponds to spinodal crossings in the negative-$H$ region
(not included in Fig.\ \protect\ref{ASFL_A7horn}).
The loops were obtained directly from the joint probability density, $P(m,m_s)$.
The rounding of $\langle |m_s| \rangle$ near the crossing of the critical line is a finite-size effect.
}\label{ASFL_A7_hysteresis}
\end{figure}

The phase diagram for $A=7$ is well suited for comparison with
a number of experimental results for SC materials that
show asymmetric, two-step thermal hysteresis loops
\cite{KOPP82,ZELE85,PETR87,BOUS92,JAKO92,REAL92,BOIN94,BOLV96,CHER04,BONN08,PILL12,BURO10,LIN12,KLEI14,CHER03,HUBY04,SHAT15,BROO15}.
Such a two-step loop, obtained directly from the joint probability density, $P(m,m_s)$,
along a path between $(H,T) = (1.0,2.5)$ and
$(-1.5,1.75)$ in Fig.\ \protect\ref{ASFL_A7horn} is shown in Fig.\ \ref{ASFL_A7_hysteresis}.
This path corresponds to the parameters $\ln r=20/3$ and $D=44/3$ in Eq.~(\ref{eq:heff}).
The narrow high-temperature loop corresponds to the crossings of the spinodal lines in
the horn region, while the wide
low-temperature loop corresponds to crossings of the spinodals in the negative-$H$ region.
In order to calculate these hysteresis loops, at each point along the hysteresis path
we first located the local maximum in $F(m)$ that separates the two phases. Then,
$\langle m \rangle$ and  $\langle |m_{s}| \rangle$ were obtained by summing over
$P(m,m_{s})$ as described in Sec.\ \ref{sec:maths_phase_diagram}.
Although we do not show other examples of hysteresis loops here, we
emphasize that our macroscopically constrained
WL method enables the calculation of such loops for any value of $A$ and any choice of
hysteresis path, solely based on the DOS data for the pure Ising antiferromagnet, without any
further MC simulations.
The hysteresis loop shown here is fully consistent with the one obtained by importance-sampling
MC simulations for the same parameters in Ref.\ \cite{PhysRevB.93.064109} \cite{DD}.
The only significant differences are the slopes of the $\langle |m_s| \rangle$ curve where the
path crosses the critical line, which in both cases are due to finite-size effects.
On the other hand, finite-size effects in the positions of the spinodals are negligible, as discussed in
Appendix \ref{sec:finite_size_error}.

\begin{figure}
\includegraphics[width=0.5\textwidth]{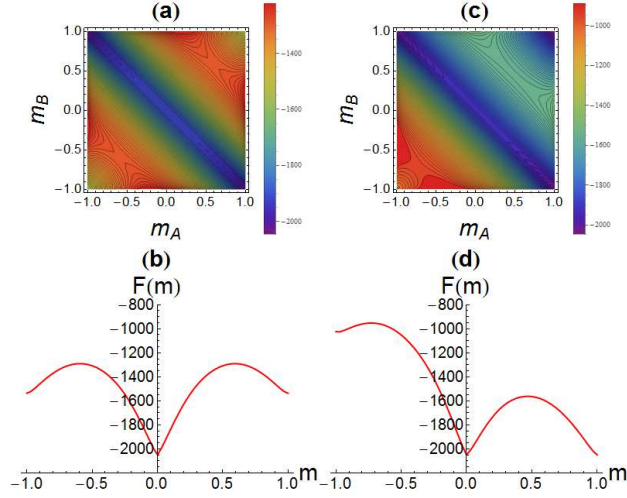}
\caption{Free-energy contour, $F(m_{A},m_{B})$ (Eq.~(\ref{def:Free_energy_pt})), and free energies, $F(m)$, for phase points with metastable FM phases for $A=7$ and $T=1$. (c)-(d) show $H=0.5005$ which is a point
lying on the coexistence line between the stable FM+ phase and the stable AFM phase. The small drop near $m=-1$ ($(m_{A},m_{B})=(-1,-1)$) indicates the presence of the metastable FM$-$ phase.  (a)-(b) show a point at the same $T$ and at $H=0$, as marked in Fig.~\ref{ASFL_phase_diagram_A6and7and8}(b). Here, AFM is stable and the two FM phases are equally metastable.
}\label{ASFL_A7_FM_negative}
\end{figure}

The phase diagrams for $A=6,7,8$ in Fig.~\ref{ASFL_phase_diagram_A6and7and8}  show several
additional, noteworthy features.
First, the phase diagrams shown are symmetric about the $T$ axis, with an exchange between FM$+$ and FM$-$. This is because the FM$+$ spinodal line is just touching the $T$ axis at $T=0$ for $A=4$ (Fig.~\ref{ASFL_phase_diagram_A1and3and4}) (c). Further increases of $A$ beyond $4$ will make a FM$-$ spinodal line show up in the positive $H$ region. Thus,
the strong $AM^{2}/(2L^{2})$ causes a FM$-$ metastable region to appear in the positive $H$ field region. Figure \ref{ASFL_A7_FM_negative}(c)-(d) illustrate the case of a point lying on the coexistence line between the FM$+$ phase and AFM phases, inside the FM$-$ metastable region. The small drop in the free energy in (d) near $m=-1$ indicates the metastable FM$-$ phase. Figure \ref{ASFL_A7_FM_negative}(a)-(b) illustrate a point at $H=0$ and at a low $T$, where both FM phases are metastable.

Second, observe that the coexistence lines turn toward stronger $|H|$ when approaching the mean-field critical points (Fig.~\ref{ASFL_phase_diagram_A6and7and8}). This is  because the disordered phase is more favorable than the FM phases at high $T$, so a stronger $|H|$ field is required to balance this effect.

Third, when $A$ increases, the mean-field critical temperature also increases, which makes the area of the horn region increase. This is because the ferromagnetic effects increase with $A$ according to the Hamiltonian (\ref{def_Hamiltonian_2D_Ising-ASFL}), so a stronger disordering effect (higher temperature) is required to balance it.

\begin{figure}
\includegraphics[width=1\textwidth]{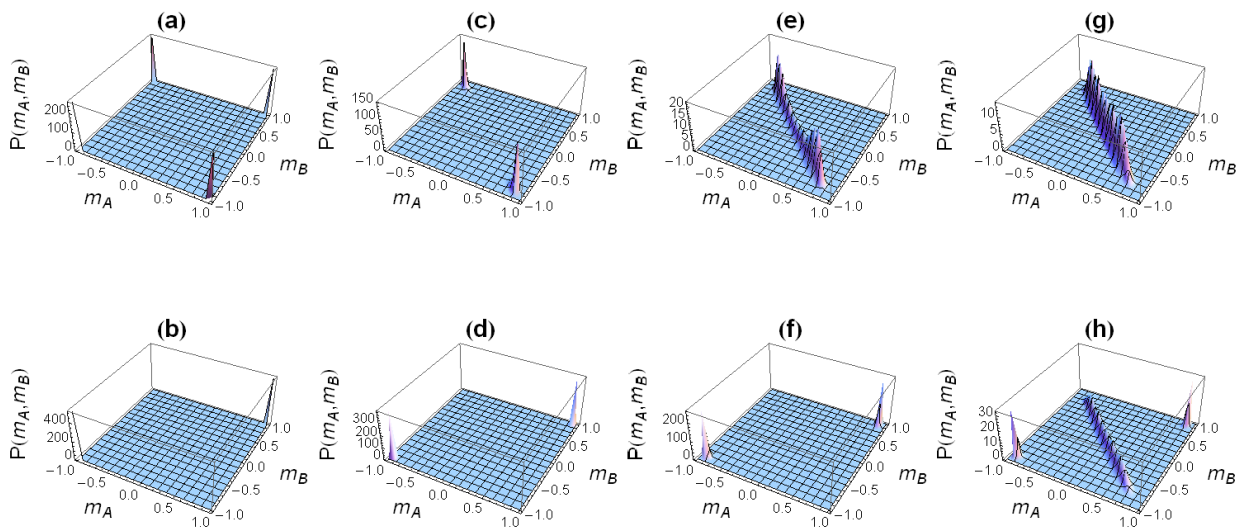}
\caption{Comparing joint probability densities for $A=8$ (first row) and $A=8.1$ (second row)
for $H=0$.
 (a)-(b) show $T=0.5$, (c)-(d) show $T=2.1$, (e)-(f) show $T=2.33$, and (g)-(h) show $T=2.38$.
 These points are marked in Figs.~\ref{ASFL_phase_diagram_A6and7and8}(c), ~\ref{ASFL_A89horn}(a), and ~\ref{ASFL_A89horn}(b).
(a) shows that all the AFM phases and the FM$\pm$ phases are stable phases for $A=8$ at low $T$ and $H=0$. (b) shows that when $A$ increases to $8.1$,
the two FM phases dominate and become the only stable phases. The two AFM phases become metastable, but are too weak to be observed in (b). (c) shows that increasing the temperature for $A=8$ makes the AFM phases become stable, and the FM$\pm$ phases become metastable. When $A$ is increased to 8.1 as in (d), the stable phases and metastable phases exchange. Similarly, (e)  shows that increasing the temperature to a point above the critical line makes the disordered phase become stable and the FM$\pm$ phases  metastable. Again, when $A$ changes to 8.1 in (f), the metastable and stable phases exchange. (g) shows that when $T$ is high enough, both $A=8$ and 8.1 will have the disordered phase as the stable phase, but the metastable FM$\pm$ phases are still visible for $A=8.1$ as shown in (h).
}\label{ASFL_A8vs8_1H0}
\end{figure}

\begin{figure}
\includegraphics[width=1\textwidth]{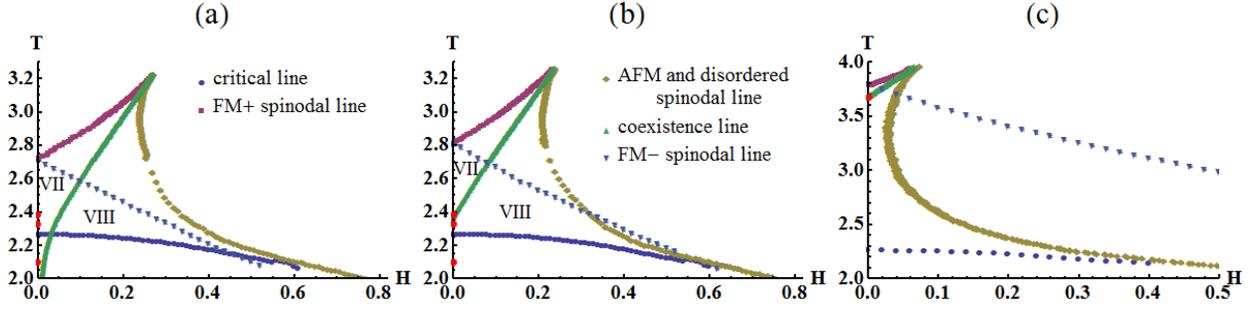}
\caption{Horn regions of the phase diagrams for (a) $A=8$, (b) $A=8.1$, and (c) A=9. Comparing with the phase diagrams for $A=6,7,$ and $8$, increasing $A$ makes the coexistence line move toward lower $|H|$, and also makes the mean-field critical point move to higher $T$. Moreover, the disordered spinodal line and the FM$-$ spinodal line intersect on the $T$ axis at a higher $T$. Note that for $A=8$, the FM$-$ spinodal line is above the critical line, which creates new metastable regions:
region VII, which is stable in the disordered phase, and metastable in both FM$\pm$ phases, and region VIII, which is stable in the FM$+$ phase, and metastable in both the FM$-$ and disordered phases.  The red dots mark the points that are discussed in Fig.~\ref{ASFL_A8vs8_1H0}. A broader view for $A=9$ is shown in Fig.~\ref{ASFL_phase_diagram_A9_9_5_11}(a).
}\label{ASFL_A89horn}
\end{figure}

Fourth, the coexistence line moves toward lower $|H|$ as $A$ increases, which makes the stable AFM region shrink and the stable FM regions grow.
This is because strong $-AM^{2}/(2L^{2})$ stabilizes the ferromagnetic phases at lower $|H|$. The coexistence line for $A=8$ is at $H=0$ for low $T$ (Fig.~\ref{ASFL_phase_diagram_A6and7and8}(c)). In that case, the two AFM phases and the two FM phases are equally probable as shown in Fig.~\ref{ASFL_A8vs8_1H0}(a). When $T$ increases to a high enough value, disorder effects start to show up, making $|m_{s}|$ decrease from 1 (Fig.~\ref{ASFL_A8vs8_1H0}(c)). At low $|H|$ and high $T$, the disordered phase is preferred over the ferromagnetic phase. This effect starts to show up before reaching the critical temperature, making the coexistence line turn away from $H=0$  before it crosses the critical line, as shown in Fig.~\ref{ASFL_A89horn}(a).

Fifth, the FM$-$ spinodal line continues moving toward higher $T$ when $A$ increases as the ferromagnetic phase is getting stronger.
At $A=8$, the FM$-$ spinodal line has moved above the critical line. This produces a region (Fig.~\ref{ASFL_A89horn}(a)) that is stable in the FM$+$ phase, and metastable in both the FM$-$ and disordered phases (region VIII), and another region that is stable in the disordered phase, and metastable in both FM$\pm$ phases (region VII).

Observe from Figs.~\ref{ASFL_A7horn} and \ref{ASFL_A89horn}(a) that the coexistence line makes a relatively large bend at the critical end-point. This is because
after passing through this point, the AFM phase changes to the disordered phase, which is favored at high temperature, making the coexistence line have a smaller slope. Therefore, a relatively large bend in the coexistence line is found at the critical end-point. This agrees with the previously observed result that $d^{2}H/dT^{2}$ along the coexistence line reaches a maximum at the critical end-point \cite{PhysRevLett.78.1488,PhysRevE.55.6624,PhysRevE.75.061108}.
Note that the location of the coexistence lines given by  Rikvold \textit{et al.}'s
Ref.\ \cite{PhysRevB.93.064109} is different from the current result for $A=7$.
The former result may be due to incomplete ergodic sampling by
the mixed-start importance-sampling MC
method used in that work to locate the coexistence lines. This might also
affect experimental attempts to accurately detect phase coexistence.
Further analysis of the discrepancy between the importance-sampling MC using the mixed-start method and the present method in locating coexistence lines is in progress \cite{CHAN17b}.

\subsection{Transitional long-range interaction strength $A=8.1,9$}
From the ground-state analysis in Ref.~\cite{PhysRevB.93.064109}, $A=8$ is the dividing line for the stable phase at $T=0$. For $A>8$, the stable phase at $T=0$, $H>0$ can only be the FM$+$ phase.
Figures \ref{ASFL_A8vs8_1H0}(a)-(d) show that increasing $A$ from $8$ to $8.1$ makes the FM phases overtake the AFM phases and become the stable phases below the critical line.
The Bragg-Williams mean-field approximation \cite{BRAG34,BRAG35} also suggests that phase diagrams having $A>8$ belong to the same group (large long-range interaction group) and
possess the same nature \cite{PhysRevB.93.064109}. While Ref.~\cite{PhysRevB.93.064109} has already pointed out that the Bragg-Williams mean-field approximation fails in predicting the existence of the horn regions (Figs.~\ref{ASFL_A7horn} and \ref{ASFL_A89horn}), here we find that the existence of the horn region induces a range of transitional long-range interaction strengths, between the medium long-range interaction and the strong long-range interaction. $A=8.1$ (Fig.~\ref{ASFL_A89horn}(b))  and $A=9$ (Fig.~\ref{ASFL_A89horn}(c)) belong to this range.

    In this transitional range of $A$, we notice several things. First, the coexistence lines still exist, but the FM phases have pushed them to meet the $T$ axis at high temperatures, and this intercept temperature increases with $A$ (Fig.~\ref{ASFL_A89horn}).
Second, while for $A=8$ and when $T$ is low, the AFM phases and the FM$\pm$ phases are equally stable along the $T$ axis (Fig.~\ref{ASFL_A8vs8_1H0}(a)). Increasing
 $A$ makes the FM$\pm$ phases overtake the AFM phases along the $T$ axis.
Figure~\ref{ASFL_A8vs8_1H0} demonstrates this by comparing four points on the $T$ axis for $A=8$ and $8.1$.
Third, the FM phases push the two spinodal lines originating from the mean-field critical point toward $H=0$.
As a result, at around $A=9$ (Figs.~\ref{ASFL_A89horn}(c)
and~\ref{ASFL_phase_diagram_A9_9_5_11}(a)), the disordered spinodal lines nearly touch the $T$ axis before the two mean-field critical points from the $\pm H$ side of the phase diagram coalesce at even higher $A$.

 \begin{figure}
\includegraphics[width=0.5\textwidth]{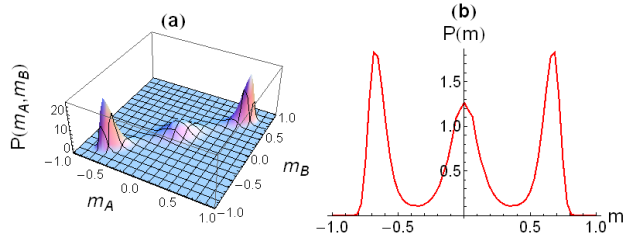}
\caption{Data for $(T,H,A)=(3.67,0,9)$, a point on the $T$ axis, close to the coexistence line for $A=9$ (marked in Fig.~\ref{ASFL_A89horn}(c)). The three peaks in $P(m)$, which correspond to the FM$\pm$ phases and the disordered phase have similar areas, and are connected with each other. When the system size is increased, the three peaks should become sharper and the connecting bridge should disappear.
}\label{ASFL_A9coexist}
\end{figure}

While Fig.~\ref{ASFL_A8vs8_1H0}(h) shows a point close to the coexistence line for $A=8.1$, which has the disordered phase spread to the two AFM corners without connecting to the two FM$\pm$ peaks, Fig.~\ref{ASFL_A9coexist} shows  a point close to the coexistence line for $A=9$, which has the disordered phase connected to the  two FM$\pm$ peaks. The connecting bridge should disappear and the three peaks should become sharper, as the system size is increased.

\subsection{Strong long-range interaction, $A=9.5,11$}
\label{sec:StrongA}
\begin{figure}
\includegraphics[width=1\textwidth]{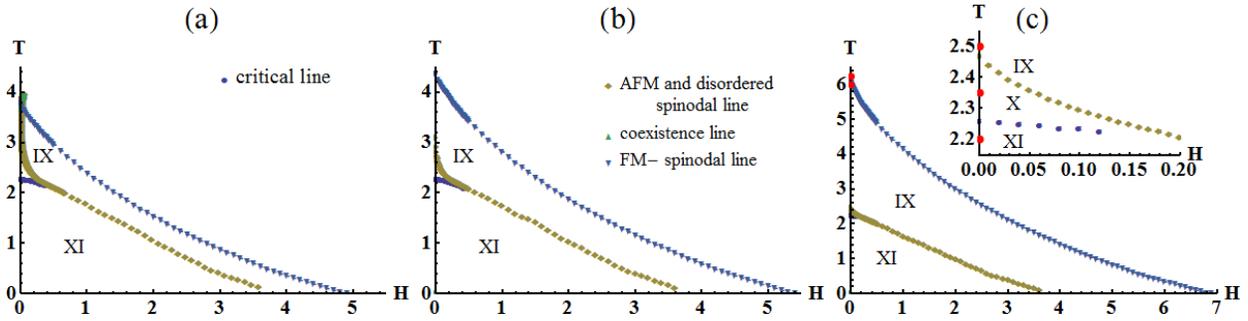}
\caption{Phase diagrams for (a) $A=9$, (b) $A=9.5$, and (c) A=11. Region IX has stable disordered/FM$+$ phase and metastable disordered/FM$-$ phase, region XI has stable FM$+$ phase and both the FM$-$ phase and the AFM phase are metastable. Region X lies near the $T$ axis between these two regions (see the inset in (c) for $A=11$). It corresponds to a  stable FM$+$ phase with both the FM$-$ phase and the disordered phase metastable.  (a) In a transitional range of long-range interaction strength, the FM phases push the disordered spinodal lines toward $H=0$, so these lines nearly touch the $T$ axis before the two mean-field critical points from the $\pm H$ side of the phase diagram coalesce. (The very small remaining horn region is shown in detail in Fig.~\ref{ASFL_A89horn}(c).) (b) and (c) correspond to strong long-range interactions.  The two mean-field critical points from the $\pm H$ side of the phase diagram have coalesced into one mean-field critical point at $H=0$, with a critical temperature that increases with $A$. The AFM and disordered spinodal lines merge with the critical line as $A$ increases. The red dots in (c) and in the inset mark phase points studied in the next few figures.
}\label{ASFL_phase_diagram_A9_9_5_11}
\end{figure}

\begin{figure}
\includegraphics[width=0.5\textwidth]{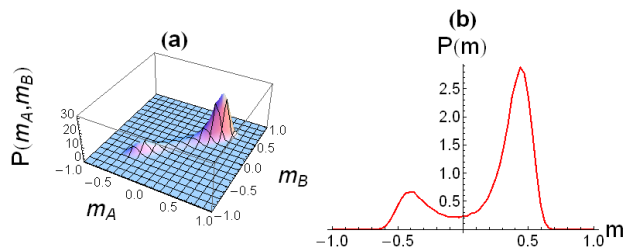}
\caption{Data for $(T,H,A)=(6,0.01,11)$, a point in region IX of Fig.~\ref{ASFL_phase_diagram_A9_9_5_11}(c).
 The marginal probability density $P(m)$ has two peaks, and the system has a stable
disordered/FM$+$ phase and a metastable disordered/FM$-$ phase.
See further discussion in Sec.~\ref{sec:StrongA}.
}\label{ASFL_A11_crossover}
\end{figure}

\begin{figure}
\includegraphics[width=0.5\textwidth]{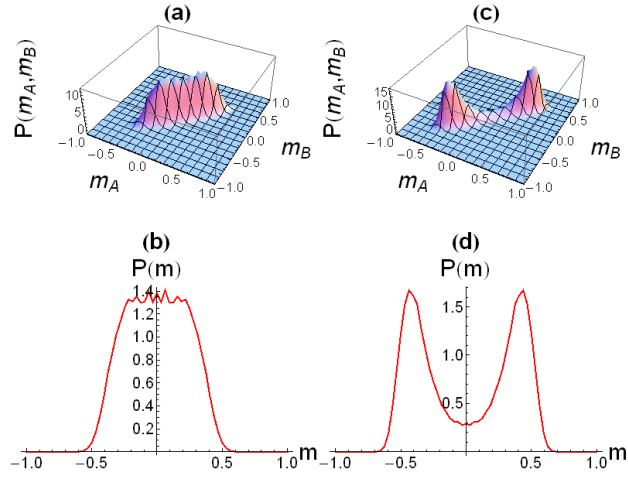}
\caption{Joint probability density, $P(m_{A},m_{B})$, and marginal probability density, $P(m)$, for $A=11$, (a)-(b) at the mean-field critical point, $(H,T)=(0,6.2239)$, (c)-(d) at a point slightly below the mean-field critical point, $(H,T)=(0,6)$ in region IX. These two points are marked in Fig.~\ref{ASFL_phase_diagram_A9_9_5_11}(c). Note that there is a continuous crossover between the disordered phase and the FM$\pm$ phase, so it is natural that the marginal probability density in (d) has big peaks at a value of $|m|$ that is smaller than 0.5, and we regard it as stable disordered/FM$\pm$ phases.
}\label{ASFL_A11_meanfield_cp}
\end{figure}

\begin{figure}
\includegraphics[width=0.5\textwidth]{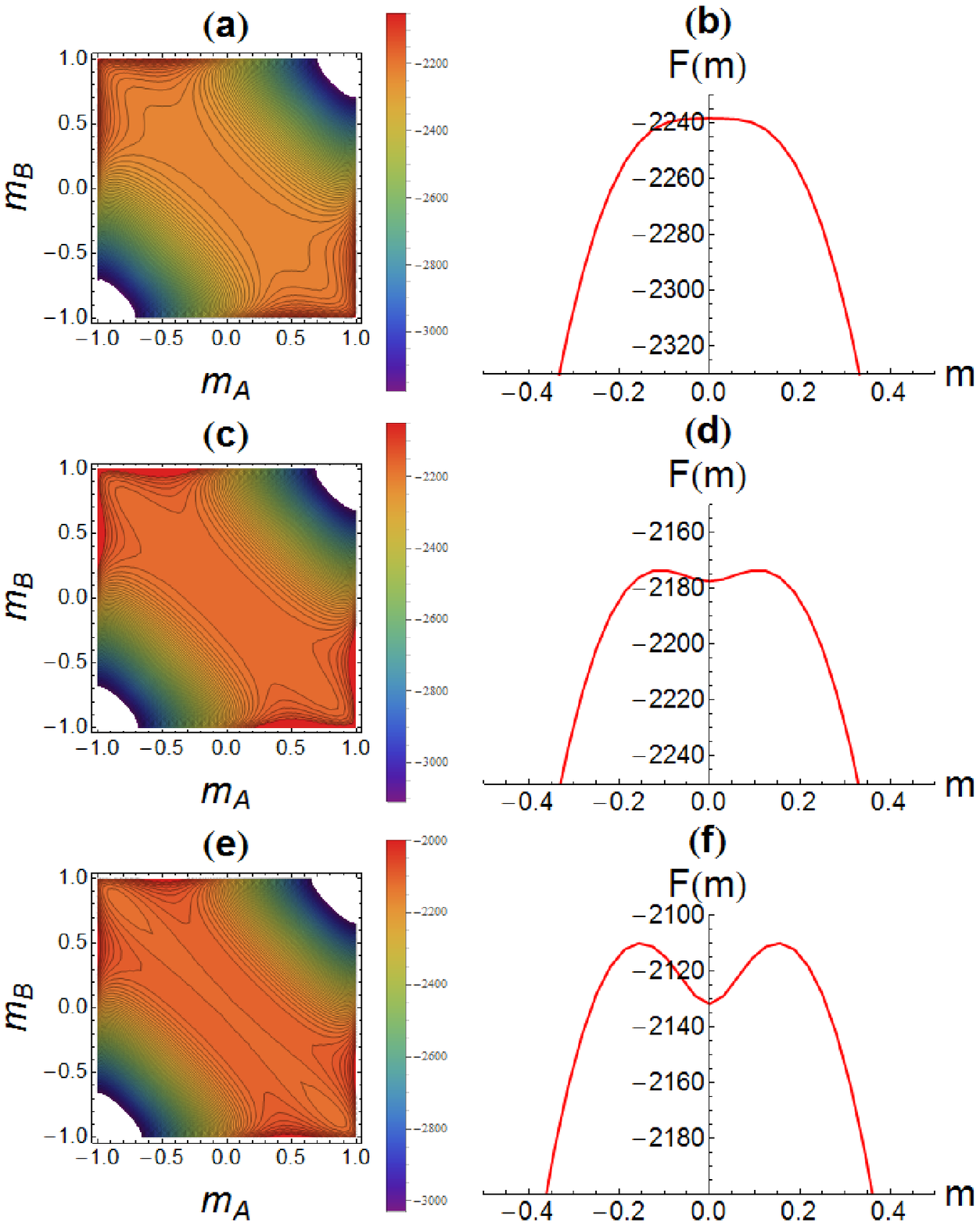}
\caption{Free energy with $A=11$ for three different points lying in the three different regions as marked in the inset of Fig.~\ref{ASFL_phase_diagram_A9_9_5_11}(c). As the three points are lying on the $T$-axis,  both the FM$\pm$ phases are stable. (a)-(b) show $(H,T)=(0,2.5)$ in region IX, which has no metastable phase, the stable phases are the FM$\pm$ which have no disordered properties, in contrast to the situation  near the mean-field critical point shown in Fig.~\ref{ASFL_A11_meanfield_cp}(d). (c)-(d) show  $(H,T)=(0,2.35)$ in  region X, which has metastable disordered phase. (e)-(f) show  $(H,T)=(0,2.2)$ in region XI, which has metastable AFM phase.
}\label{ASFL_A11_metastable}
\end{figure}

When the long-range interaction is sufficiently strong, the two mean-field critical points in the $\pm H$  horn regions will coalesce into one critical point as shown in the phase diagrams for $A=9.5$ and $11$ in Fig.~\ref{ASFL_phase_diagram_A9_9_5_11}(b)-(c).
Above this mean-field critical temperature, the system is in a disordered phase. If we increase $H$, the system undergoes a continuous crossover from the disordered phase to the FM$+$ phase, but there is no sharp transition point. The combined mean-field critical point is also the end point of the FM$\pm$ spinodal lines.
 When $H>0$ and $T$ is below the FM$-$ spinodal line (region IX in Fig.~\ref{ASFL_phase_diagram_A9_9_5_11}), the marginal probability density $P(m)$ has two peaks, and the system has a stable disordered/FM$+$ phase and a metastable disordered/FM$-$ phase  (Fig.~\ref{ASFL_A11_crossover}).
As there is a continuous crossover between the disordered phase and the FM$+$ phase above $T_{c}$, it is natural that near the mean-field critical point, the marginal probability density has a large peak at a value of $|m|$ that is smaller than 0.5. Moreover, the metastable phase can show very strong disordered properties, so we consider the metastable phase below the FM$-$ spinodal line to be a disordered/FM$-$ phase. The topology of the phase diagrams for $A=9.5$ and $11$ is the same as found for $A=10$ in Ref.~\cite{PhysRevB.93.064109}.

Fig.~\ref{ASFL_A11_meanfield_cp}(a)-(b) show probability densities at the coalesced mean-field critical point. It is found by extrapolation of the FM$-$ spinodal line and Eq.~(\ref{spinodal_vs_temp}).
 Note that, as the critical point is in the mean-field universality class, at $T=6$, which is below the critical point for $A=11$ as shown in Fig.~\ref{ASFL_A11_meanfield_cp}(c)-(d), we regard
it as having stable FM$\pm$ phases, connected by fluctuations resembling the disordered phase. However, we do not regard the system as having a metastable disordered phase.
The fluctuation connection has disappeared at around $T=5.5$. At $T=2.5$ as shown in
Fig.~\ref{ASFL_A11_metastable}(a)-(b), the system is close to the AFM and disordered spinodal line, the free energy in (b) shows a flat maximum around $m=0$.
Figure \ref{ASFL_A11_metastable}(c)-(d) shows the case at $T=2.35$ for $A=11$,
which is a point in region X in the phase diagram of
Fig.~\ref{ASFL_phase_diagram_A9_9_5_11}(c).
The free-energy contour  and the free-energy drop near $m=0$ indicate the existence of the metastable disordered phase.
 Further reduction in $T$ below the critical line brings the system to the stable FM$+$ phase with metastable AFM phases, i.e. region XI in the phase diagram of
Fig.~\ref{ASFL_phase_diagram_A9_9_5_11}(c),
as shown in Fig.~\ref{ASFL_A11_metastable}(e)-(f) for $T=2.2$. The free-energy contour, and the drop in free energy near $m=0$, indicate the existence of the metastable AFM phases.
As $A$ increases (Fig.~\ref{ASFL_phase_diagram_A9_9_5_11}), the disordered/AFM spinodal
line merges with the critical line. We expect region X, the disordered metastable phase region, to disappear when $A$ becomes very large.

\section{Summary and Conclusion}\label{sec:conclusion}
In this paper we have presented detailed phase diagrams, free-energy landscapes, and order-parameter distributions for a model SC material with antiferromagnetic-like nearest-neighbor and ferromagnetic-like long-range interactions \cite{ PhysRevB.93.064109}, covering a wide range of temperatures $T$, fields $H$, and long-range interaction strengths $A$. This was accomplished with a relatively modest computational effort by a recently developed, Macroscopically
Constrained WL method for systems with multiple order parameters \cite{ PhysRevE.95.053302}. The method produces DOS for given values of the system energy $E$, magnetization $m$, and staggered magnetization $m_s$ for a square-lattice Ising antiferromagnet ({\it i.e.\/}, $A=0$) in zero field. The DOS for arbitrary values of $H$ and $A$ are then found by a simple transformation of $E$ [Eq.~(\ref{def_shift_E_DOS})], without the need for additional simulations. From the transformed DOS, we obtain free-energy landscapes and $(H,T)$ phase diagrams, including metastable regions important to applications of SC materials \cite{OHKO11,CHAK14,KAHN98,LINA12}.
Topologically different phase diagrams are obtained, depending on the strength of $A$. For $A=0$, the numerically well-known phase diagram for the square-lattice antiferromagnet is recovered (Fig.\ \ref {ASFL_phase_diagram_criticalline}).

For weak long-range interactions, $0 < A \lesssim 4$, the high-temperature critical line terminates in a tricritical point at a nonzero temperature, from which sharp spinodal lines marking the extent of metastable phase regions extend to $T=0$ (Fig.\ \ref{ASFL_phase_diagram_A1and3and4}).
In this parameter range, the phase diagram is topologically identical to what is predicted by a simple Bragg-Williams mean-field approximation as discussed in Ref.\ \cite{ PhysRevE.95.053302}.

At a value of $A$ between 4 and 6 (which we have not attempted to determine accurately), the tricritical point decomposes into a critical end-point and a mean-field critical point at a higher temperature. The resulting horn structure of the phase diagram, which is not seen in simple Bragg-Williams mean-field calculations, is illustrated in Fig.~\ref{ASFL_phase_diagram_A6and7and8}
for the intermediate interaction strengths, $A=6$, 7, and 8. The phase diagram obtained for $A=7$
(Fig.~\ref{ASFL_A7horn})
is in excellent agreement with that obtained by computationally intensive importance-sampling MC simulations in Ref.\ \cite{ PhysRevB.93.064109}.
The only clear difference is the shape of the AFM/FM coexistence lines. A detailed investigation of this issue is in progress \cite{CHAN17b}.
(Very recently, horn regions and asymmetric, two-step hysteresis loops,
analogous to those seen in the model studied here, have also been observed for
a model with antiferromagnetic-like nearest-neighbor
interactions and genuine elastic interactions \cite{NISH17}.)
The horn structure gives rise to asymmetric, two-step
hysteresis loops (see example in Fig.\ \ref{ASFL_A7_hysteresis}) that are similar to
experimental observations in several different SC materials
\cite{KOPP82,ZELE85,PETR87,BOUS92,JAKO92,REAL92,BOIN94,BOLV96,CHER04,BONN08,PILL12,BURO10,LIN12,KLEI14,CHER03,HUBY04,SHAT15,BROO15}.

For $A>8$, the AFM phase is no longer a possible ground state of the model. In the transitional region, $8 < A \lesssim 9$, the horn region shrinks as shown in Fig.\ \ref{ASFL_A89horn}, until the two mean-field critical points coalesce into a single critical point at $H=0$ for a value of $A$ somewhere between 9 and 9.5. (This value we also have not attempted to determine accurately.) To our knowledge, this regime of transitional interaction strengths has not been investigated before.
Phase diagrams for the strong-interaction case, represented by $A=9.5$ and 11, are shown in Fig.\ \ref{ASFL_phase_diagram_A9_9_5_11}. These are topologically identical to the one shown for $A=10$ in Ref.\ \cite{ PhysRevB.93.064109}.
We believe our results can contribute to the interpretation of the fascinating phase diagrams and hysteresis loops observed in many SC materials and other systems with competing
short- and long-range interactions.

\section*{ACKNOWLEDGMENTS}\label{sec:acknowledgement}

 The Ising-ASFL model was first proposed by Seiji Miyashita,
and we thank him for useful discussions.
The simulations were performed at the Florida State University High Performance Computing
Center.
This work was supported in part by U.S.\ National Science Foundation grant No.\ DMR-1104829.

\appendix

\section{Finite-size effects and error estimates}
\label{sec:finite_size_error}
\begin{figure}
\includegraphics[width=0.5\textwidth]{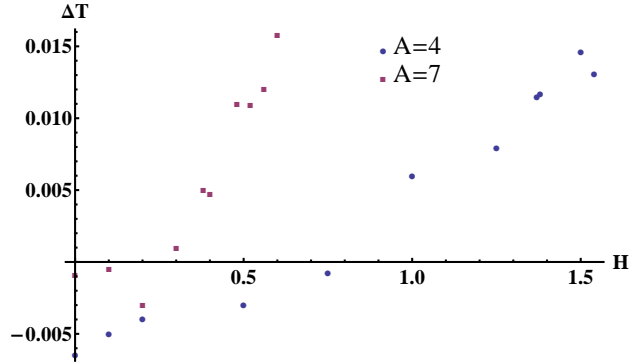}
\caption{Comparison of points on the critical lines for $A=4$ and $A=7$
obtained in this paper with $L=32$, and those obtained through importance-sampling
MC with $L \le 1024$ in Ref.~\cite{PhysRevB.93.064109}. For each
value of $H$, $\Delta T>0$ means
that Ref.~\cite{PhysRevB.93.064109} locates the critical point at a higher temperature than
obtained in the present paper. As $H$ increases,
$\Delta T$ increases from negative to positive. However, even near the end of the critical line, this error is less than the symbol size in all the phase diagrams shown in the present paper,
except Fig.\ \ref{ASFL_A7horn}.
}\label{ASFL_finite_size_error}
\end{figure}

The questions of finite system sizes and error estimates are intimately connected, and it is
reasonable to ask whether the system size of $L=32$ that we use here is sufficient to ensure
reliable results. The fourth-order Binder cumulant presumably leads to cancellation of leading
corrections to scaling \cite{PhysRevLett.47.693} and is a remarkably
accurate method to locate critical points.
The most general way to utilize the method is to look for the crossings between plots of cumulant
vs temperature or field for different system sizes. However, the model studied here
fulfills all the symmetry requirements to yield a fixed-point value
of $0.6106924(16)$ \cite{0305-4470-26-2-009,PhysRevE.70.056136,0305-4470-38-44-L03,Salas2000}. As a consequence, it is possible to obtain good estimates of critical points as
the phase points where the cumulant is near
this value for a single system size, as we have done here.
This is demonstrated in
Fig.~\ref{ASFL_finite_size_error}, where we compare critical lines obtained here using the macroscopically constrained WL method with $L=32$, with those obtained
in Ref.~\cite{PhysRevB.93.064109} by importance-sampling MC using the standard method
of cumulant crossings for $L \le 1024$. The differences are indeed
very small, and although they are included as error bars in all the phase diagrams shown
in this paper,
they only exceed the symbol size in the lower right quadrant of the
enlarged view of the horn region for $A=7$, shown in Fig.\ \ref{ASFL_A7horn}.
The finite-size effects are even smaller for the spinodal lines (not shown here),
and again the error bars obtained
from the differences with the results of Ref.~\cite{PhysRevB.93.064109} are only visible in
Fig.\ \ref{ASFL_A7horn}.
Statistical errors were reduced below the level of the finite-size effects by averaging the DOS over
ten independent macroscopically constrained WL simulations as described in Appendix C
of Ref.\ \cite{PhysRevE.95.053302}.



\begin{thebibliography}{72}
\bibitem{DD}
We note that the parameter $D$ used for the hysteresis path and loop shown in
Figs.\ 5(a) and 8 of Ref.\ \cite{PhysRevB.93.064109},
respectively, is there misprinted as 18. The correct value
is 44/3, the same as used in the present paper.
\end{thebibliography}

\ifx \manfnt \undefined \font\manfnt=logo10 \fi

\end{document}